\documentclass{emulateapj}

\usepackage{color}
\usepackage{graphicx,epstopdf}
\usepackage{bmpsize}
\usepackage{enumerate}

\newcommand{\elem}[1]{{\scriptsize~{#1}}}

\newcommand{\phot}{~photons~cm$^{-2}$~s$^{-1}$~sr$^{-1}$}

\newcommand{\comRef}[1]{{\bf \textcolor{black}{#1}}}
\newcommand{\PS}{power spectrum}
\newcommand{\ACT}{\emph{ACT}}
\newcommand{\SPT}{\emph{SPT}}
\newcommand{\Planck}{\emph{Planck}}

\shorttitle{X-ray and Sunyaev-Zeldovich properties of the WHIM}
\shortauthors{Ursino, E., Galeazzi, M., and Huffenberger, K.}

\begin{document}
\title{X-ray and Sunyaev-Zeldovich properties of the Warm-Hot Intergalactic Medium}


\author{Ursino, E., Galeazzi, M.\altaffilmark{1}, and Huffenberger, K.\altaffilmark{2}}
\affil{Physics Department, University of Miami, Coral Gables, FL 33155}
\altaffiltext{1}{corresponding author, galeazzi@physics.miami.edu}
\altaffiltext{2}{current address, Physics Department, Florida State University, Tallahassee, FL 32306}


\begin{abstract}

We use numerical simulations to predict the soft X-ray ([0.4-0.6] keV) and Sunyaev-Zeldovich signal (at 150~GHz) from the large scale structure in the Universe and then compute 2-point statistics to study the spatial distribution and time evolution of the signals. The average X-ray signal predicted for the WHIM is in good agreement with observational constraints that set it at about 10\% of the total Diffuse X-ray Background. The characteristic angle computed with the Autocorrelation Function is of the order of some arcminutes and becomes smaller at higher redshift. The \PS\ peak of the SZ due to the WHIM is at $l\sim10000$ and has amplitude of $\sim0.2$~$\mu$K$^2$, about one order of magnitude below the signal measured with telescopes like \Planck, \ACT, and \SPT. Even if the high-redshift WHIM signal is too weak to be detected using X-rays only, the small-scale correlation between X-ray and SZ maps is dominated by the high-redshift WHIM. This makes the analysis of the SZ signal in support of X-rays a promising tool to study the early time WHIM.

\end{abstract}



\section{Introduction}
\label{introduction}
Theoretical and observational investigations of the baryon density at high and low redshift show discrepancies. At high redshift ($z>2$) we see that $\Omega_b\sim 0.045$ \citep{Weinberg97, Rauch98, BurTyt98, Kirkman03, Bennett03, Komatsu11}, while at low redshift the observed $\Omega_b$, despite the recent improvement in X-ray and FUV observations, is up to 40\%  smaller \citep{Fukugita98, Shull12} and about $30\%$ of the baryons are still unaccounted for. A possible solution to the problem of the missing baryons is that they reside in the Warm-Hot Intergalactic Medium (WHIM), a filamentary gas with $10^5<T<10^7$~K and density smaller than 1000 times the mean baryonic density \citep{CenOst99}. According to hydrodynamical simulations, at present time about half of the baryons are in the WHIM \citep{CenOst99,Borgani04,CenOst06,Tornatore10}, in the form of a highly ionized plasma that emits mostly in the UV and low-energy X-ray bands. 

While the WHIM contains such a large fraction of baryons, it is however hard to detect its signal as there are other sources of X-rays, like the Solar Wind Charge Exchange (SWCX), the Local Hot Bubble (LB), the Galactic Halo (GH), clusters of galaxies, and unresolved point sources that dominate the soft X-ray Universe. In comparison, the WHIM emission contributes to $10-15\%$ of the total Diffuse X-ray Background (DXB) \citep{Phillips01, Galeazzi09}. Clusters of galaxies can be identified and removed from maps, so that they do not compete with the WHIM signal. Unresolved point sources, as well, provide a well characterized power law spectral signature that can be easily modeled. GH and LB, instead, are thermal plasmas at temperatures comparable with the WHIM but with much stronger contribution to X-rays. The SWCX has only line emission, but the lines correspond to the same characteristic emission lines (most importantly O\elem{VII} and O\elem{VIII}) of a $\sim10^6$~K plasma like the WHIM. The WHIM emission could be still disentangled from LB, GH, and SWCX by searching for the redshifted emission lines (LB, GH, and SWCX are local). The relatively poor energy resolution ($\sim70$~eV) of present day X-ray telescopes, however, makes this approach a very hard task.

Presently, the best evidence of WHIM detection comes from absorption lines in the FUV spectra of distant AGN, although the absorbers primarily trace gas at $T<10^6$~K or even highly photoionized gas at $T\sim10^{4.5}$~K \citep{Danforth05, Danforth08, Tripp08}, and by broad Ly$\alpha$ absorbers \citep{Richter04, Danforth10}. Detections of absorption in the Soft X-rays ($E<2$~keV), where most of the WHIM is expected to emit, proved to be much more difficult and so far they are only a handful \citep{Nicastro05, Fang02, Fang07, Buote09, Zappacosta10}. These detections, however, are controversial: in some case there is debate about their statistical significance \citep{Rasmussen07}, while more in general they could actually trace the gas that surrounds galaxies \citep{Williams13}. Direct evidence of WHIM emission, on the other side, has been found in the direction of a filament connecting the two clusters A222 and A223 \citep{Werner08}. Indirect evidence comes from the analysis of the autocorrelation function of the DXB in the [0.4-0.6]~keV band \citep{Galeazzi09}. \citet{Cappelluti12} performed a similar analysis on the Chandra Deep Field South using the \PS\ instead of autocorrelation, and did not find any significant contribution from the WHIM. However, it has to be noted that they studied the signal in the [0.5-2.0]~keV energy band, where the WHIM emission is almost negligible compared to galaxy clusters.

Models based on hydrodynamical simulations \citep{Borgani04} predict that the WHIM filaments show a characteristic signal at angles of a few arcminutes that can be used to disentangle the WHIM from the other components of the DXB signal \citep{Ursino11}. However, while the combination of these detections seems to confirm the existence of the WHIM, the small number of samples studied and the uncertainties introduced by the competing background sources and the instrumental noise do not allow for a significant characterization of the gas in the WHIM.

In addition to absorption and emission associated with the ions in the gas filaments, it is in principle possible to detect a signal associated with the free electrons in the hot plasma in microwave data through the Sunyaev-Zeldovich effect. Both thermal and kinetic (bulk) motions of the gas contribute to the spectrum of the cosmic microwave background (CMB, \citealt{SZ70, SZ72, SZ80, SZ81}). The effect provides further information about the properties and distribution of the energetic gas in the WHIM. In this work we focus on the larger, thermal SZ effect. From now on we use the notation SZ to refer to the thermal Sunyaev-Zeldovich effect only, leaving the notation tSZ and kSZ when we clearly need to distinguish between the two effects. In the SZ effect, electrons in hot gas along a line of sight scatter low-energy photons in the microwave background. Photons scattered into the line of sight have preferentially higher energy than those scattered out, conserving the total number of photons and imprinting a CMB spectral distortion in the direction of concentrations of hot electrons in the universe. 

Observationally the SZ effect is well established by detection of the shock heated gas in galaxy clusters. The largest SZ surveys performed with the \emph{South Pole Telescope} \citep{Ruhl04, Staniszewski09, Vanderlinde10}, the \emph{Atacama Cosmology Telescope} \citep{Kosowsky03, Marriage11, Hasselfield13}, and the \Planck\ satellite \citep{Ade13a} have mapped thousands of square degrees at arcminute resolution (or several arcminutes in the case of \Planck).

Hydrodynamical simulations \citep{Roncarelli07} indicate that roughly $20\%$ of the SZ signal comes from overdense unbound objects, like the WHIM, but the signal is too low to be detected with current surveys. However, investigating the correlation between microwave and X-ray data could raise the signal above the current noise levels. Due to its distinct emission mechanism, SZ would provide additional constraints on the structure of the Intergalactic Medium (IGM). As we will see more deeply in section~\ref{Modeling_section}, the SZ effect is the line-of-sight integral of pressure. In our simulation \citep{Borgani04}, and usually for many simulations and observations (\citealt{Ostriker05} and references therein), pressure depends on density following a polytropic relation ($P \propto \rho^{1.2}$), so that the SZ temperature fluctuations go as $\rho^{1.2}$. The X-ray emission, instead, goes as $\rho^2$ (slightly modified by the cooling function). The SZ signal is therefore expected to be more sensitive to lower-density regions than the X-ray one, and the combination of the two allows a fuller study of the density distribution of the gas in the WHIM filaments.  

In this paper we use the output of the large scale hydrodynamical simulation by \citet{Borgani04} to generate X-ray and SZ maps of the WHIM, we then characterize the average signal and the 2-point statistics, we study the cross-correlation between X-rays and SZ and we analyze the redshift evolution of the signals. This way we can investigate the properties of the SZ signal from the WHIM and its combination with X-ray emission. Our previous work on the WHIM \citep{Ursino06,Ursino10, Ursino11} has shown that the filaments have typical angular scales of a few arcminutes and therefore we focused our attention to the angular range 0.3-10 arcminutes.

The paper is organized as follows: in section~\ref{simul_tools} we describe the model used for the simulations and the methods used to construct the X-ray and SZ images and we briefly review the tools we use for the statistical analysis. In section~\ref{test_model} we compare our results with previous simulations and with observations to test the robustness of our model. Then, in section~\ref{large_scale_XSZ}, we show the predictions from our simulations and we analyze the properties of the temperature fluctuations and the X-ray emission due to large scale structures. In section \ref{Redshift_Evolution} we follow the redshift evolution of the gas phases, with particular focus on the WHIM and on clusters, and we show how it affects the evolution of the SZ effect and X-ray emission. Finally, in section \ref{conclusion}, we summarize and discuss the results. 

\section{Methods}
\label{simul_tools}

\subsection{Modeling the X-ray and SZ universe}
\label{Modeling_section}

Here we give only a brief description of the hydrodynamical simulation adopted and the procedures we used to generate X-ray maps, referring the reader to \citet{Borgani04} and \citet{Ursino10,Ursino11} for more details. Instead, we focus on the procedure we used to generate the SZ maps.

The hydrodynamical simulation was performed with the TREESPH code \emph{GADGET-2} \citep{Springel01, Springel05}, using a $\Lambda$CDM model with cosmological constant $\Omega_\Lambda= 0.7$, $\Omega_\textrm{m} = 0.3$, and a baryon density $\Omega_\textrm{b} = 0.04$, the Hubble constant is $H_0=100$~h~km~s$^{-1}$~Mpc$^{-1}$, with $\textrm{h} = 0.7$, and $\sigma_8 = 0.8$. It follows the evolution of $480^3$ dark matter and baryonic gas particles, from $z = 49$ to $z=0$. The model includes  gravity, non-radiative hydrodynamics, star formation, feedback from SNe with the effect of weak galactic outflows, radiative gas cooling and heating by a uniform, time-dependent, and photoionizing ultraviolet background. The output is a set of 102 boxes with a side of 192~h$^{-1}$~Mpc, ranging from $z=9$ to $z=0$. At the time of our work there were other hydrodynamical simulations describing dark matter, baryonic matter, and metals evolution with more advanced models \citep{Oppenheimer08,Schaye10,Tornatore10}.  Developement in the past few years, in fact, have shown that AGN feedback is a key parameter for the evolution of baryons both within clouds and at the outskirts, where clusters blend with the WHIM filaments \citep{Sijacki07, Puchwein08,  Battaglia10, Bertone10, Fabjan10, McCarthy10, Planelles14}. However the comoving box of the newer simulations contains less than 1/8 the volume of the Borgani et al. (2004) simulation, and would allow fewer than three independent fields of view.  To have statistical power, we opted for the older, less advanced, but larger simulations. Moreover, we have not observed any significant change to the WHIM predictions of the simulations, making the one by \citet{Borgani04} still significant for the proposed study. We note that, while finalizing this paper, \citet{McCarthy14} showed the results of the analysis of a very large simulation box with volume (400~$h^{-1}$~Mpc)$^3$, that includes advanced subgrid treatment. This new simulation looks as a very promising candidate for a statistical analysis of the WHIM X-ray emission and SZ properties.

In order to mitigate some of the deficiencies of the simulation we adopted an improved metallicity model (since metallicity is still by far the largest unknown constraints from observations). In previous work \citep{Ursino10}, we compared the effects of several metallicities (none of which exceeds observational constrains) and in the current work we adopted the model that is in best agreement with constraints from X-ray observations (see section \ref{Xray_test}) and Ly-$\alpha$ observations \citep{CenOst99b}. The model is also in good agreement with updated large scale surveys \citep{Pettini06, Rafelski13}.

We stacked the simulation boxes in the redshift interval from $z=0$ to $z=3$ and then projected the product of particle mass and temperature (required to generate SZ maps) and the X-ray flux from the single particles to a mesh using a 3-dimensional smoothing kernel as a weight \citep{Monaghan85}. The volume elements of the mesh have size of $1/256\times1/256$~deg$^2\times3$~Mpc~h$^{-1}$ (the goal is to create $1\times1$~deg$^2$, $256\times256$ ~pixels maps). In order to avoid that the same structure falls in a single field of view at different redshifts, we performed random rotations and shifts to the particles in each simulation cube. 

For every baryonic particle in the simulation we have density, volume, mass, and position. We used the physical parameters to generate the metallicity, the X-ray emission, and the contribution to the SZ effect.

The metallicity is randomly assigned to the particles using a probability distribution function of density and redshift following Fig.~2 of \citet{CenOst99b}. 

Using the \textit{apec } model for \textit{XSPEC}\footnote{http://heasarc.gsfc.nasa.gov/docs/xanadu/xspec/} we generated tables of the energy spectrum (with an energy resolution of 1~eV) as a function of temperature and metallicity. By interpolating on temperature and metallicity, and multiplying by $\rho^2$ we can create an X-ray spectrum for every particle. We summed on the redshifted energy bins (using the particle redshift) to compute its contribution to the $0.4-0.6$~eV band. The X-ray emission of the single particle is finally smoothed over the volume elements associated to the particle's position. This energy band is chosen because it contains the O\elem{VII} and O\elem{VIII} emission lines up to a redshift $\sim0.5$, the main trackers of the WHIM. We do not include lower energy (although it could probe higher redshifts) because most X-ray detectors have a high noise level below 0.4~keV.

To compute the temperature fluctuations of the CMB induced by the SZ effect we used the following procedure:
\begin{equation}
\label{sz_01}
\frac{\Delta T_{CMB}}{T_{CMB}}=g(x)y,
\end{equation}  
where
\begin{equation}
\label{sz_02}
y=\int \sigma_T n_e \frac{k_BT_e}{m_ec^2}dr
\end{equation}
and
\begin{equation}
\label{sz_03}
g(x)=x\; coth(x/2)-4,
\end{equation}
and
\begin{equation}
\label{sz_04}
x=\frac{h\nu}{k_BT_{CMB}},
\end{equation} 
with $\sigma_T$ the Thompson constant, $k_B$ the Boltzmann constant, $m_e$ the electron mass, $c$ the speed of light, $h$ the Plank constant, $\nu$ the radiation frequency, and $T_{CMB}$ the CMB temperature (2.725~K). The free parameters are $n_e$, the electron density, and $T_e$, the gas electron temperature (in WHIM, the electron temperature requires no relativistic corrections to the frequency dependence). In order to compute $n_eT_e$, and therefore the Compton $y$ parameter (eq.~\ref{sz_02}), we used the weighted mass$\times$temperature term of every single volume element. The ``mass-temperature'' term is obtained by computing it for each single particle, then smoothing it over the corresponding volume elements (in the same way as we did for the X-ray emission), and finally summing over all the contributions within a single volume element. 

We evaluated $g(x)$ for a frequency of 150~GHz (using equations~\ref{sz_03} and \ref{sz_04}). Finally, we multiplied the $y$ coefficient by $g(x)$ and $T_{CMB}$ and obtained the fluctuation $\Delta T$. 

By adding along the z coordinates we generated the two dimensional X-ray and SZ maps both for the full line of sight up to $z=3$ and for selected redshift interval in order to evaluate redshift evolution.  

The mapping program can also filter the simulation particles by their density and temperature. This way we could define the WHIM (gas with $10^5<T<10^7$~K and overdensity $\rho<1000\rho_b$), dense WHIM ($10^5<T<10^7$~K and $\rho>1000\rho_b$, generally associated to groups of galaxies), and clusters ($T>10^7$~K). For this work, however, we generated maps of the WHIM alone and maps of all the gas at $T>10^5$~K (WHIM, dense-WHIM, and clusters). In general, low temperature gas has low density, and therefore its X-ray emission is negligible: this is the reason why we ignored gas at low temperature. Since most of the particles in the simulation are low temperature particles, ignoring them reduces the computation time by a at least a factor 2.
From now on we will use the following notation: \emph{``WHIM''} stands for gas with $10^5<T<10^7$~K and $\rho<1000\rho_b$, \emph{``all gas''} (or \emph{``IGM''}) stands for gas at $T>10^5$~K, regardless of overdensity, and \emph{``non-WHIM''} stands for all the gas except the WHIM. Overall, we generated 10 all-gas maps and the corresponding WHIM maps.

As we will show in section~\ref{large_scale_XSZ}, we used the set of all-gas maps to estimate the reliability of our predictions, comparing them with previous works. It is worth noting that, while other works \citep{Roncarelli07} showed that there is contribution to the SZ effect from gas up to $z=7$, we have seen that the WHIM contribution to the \PS\ signal is already almost negligible at $z=3$. In previous works \citep{Ursino06, Ursino10} we have already proven 	that the bulk of the X-ray emission comes from even closer distances. Hence our choice to limit the simulation to $z=3$.  

\begin{figure*}
\plottwo{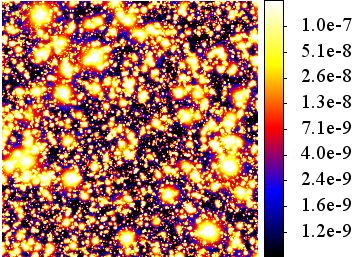}{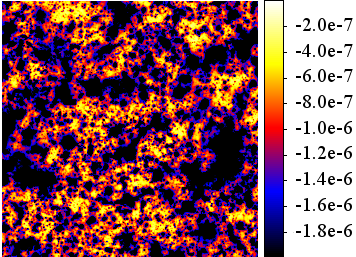}
\plottwo{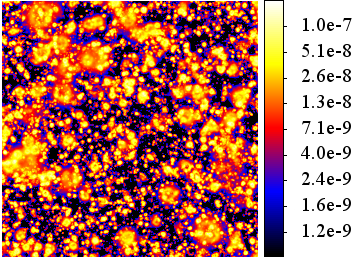}{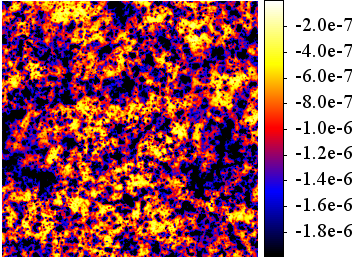}
\caption{Sample IGM (\emph{top}) and WHIM (\emph{bottom}) X-ray (\emph{left}) and SZ (\emph{right}) map at 150~GHz. X-rays are in units of photons~cm$^{-2}$~s$^{-1}$, temperature fluctuations are in K. The maps show good correlation between strong X-ray emission (brighter pixels) and larger temperature fluctuations (darkest SZ pixels). Color version on-line only.
\label{maps}}
\end{figure*}

In figure~\ref{maps} we show the WHIM X-ray (\emph{left}) and SZ (\emph{right}) maps for one field-of-view extracted from the simulation. In order to enhance the structure of the X-ray emission and the temperature fluctuations, we plotted the maps in non-linear scales. By visual inspection it is evident that they trace the same underlying structure, where regions with strong X-ray emission (brighter) have bigger negative temperature fluctuations.

\subsection{Angular Autocorrelation Function and Power Spectrum}
\label{AcF_PS}
For this investigation we need to compute the angular properties of the X-ray and SZ maps. This analysis can be performed either using the Angular Autocorrelation Function (AcF) or computing the Power Spectrum (PS). It is well-known that the AcF and the PS are related to each other via the Fourier transform, however in this work we will evaluate AcF and \PS\ independently.

Historically, the tool of choice for the analysis of microwave maps is the PS. For X-ray maps, instead, the AcF is preferred. Although much slower ($N^2$ vs. $Nlog N$ iterations, where $N$ is the number of pixels in the map), the AcF offers an easy and natural way of dealing with the large number of pixels that need to be removed because they contain point sources or very bright extended objects.
 
We adopt the following AcF estimator:
\begin{equation}
\label{AcF}
w(\theta)=\langle R(\mbox{\boldmath$n$}) R(\mbox{\boldmath$n'$})\rangle-\langle R \rangle^2,
\end{equation}
where $R(\mbox{\boldmath$n$})$ is the value  of X-ray flux or $\Delta T$ at a line of sight \emph{\textbf{n}} and $R(\mbox{\boldmath$n$})$ and $R(\mbox{\boldmath$n'$})$ are separated by $\theta$. The estimator is a variant of the one adopted by \citet{Soltan99}, the main difference with both our previous work \citep{Ursino06, Galeazzi09} and other works (i.e. \citealt{Soltan99}, \citealt{Landy93}) is that our estimator provides a dimensioned AcF, while usually the AcF is dimensionless (i.e. we do not normalize the signal to the average of the map). Using a dimensioned estimator makes it easier to control the order of magnitude of the results and it is directly comparable with the magnitude of the \PS.

The estimator of the cross-correlation between the X-ray and SZ fields is similar to eq.~\ref{AcF}:
\begin{equation}
\label{CcF}
w(\theta)=\langle X(\mbox{\boldmath$n$}) SZ(\mbox{\boldmath$n'$})\rangle-\langle X \rangle \langle SZ\rangle,
\end{equation}
where $X$ and $SZ$ clearly represent the values of the pixels in the X-ray and SZ maps.

To calculate the \PS\ we introduced the parameter $\Theta(\mbox{\boldmath$n$})=\Delta T(\mbox{\boldmath$n$})/T$. Using the flat sky approximation, in Fourier space this goes as:
\begin{equation}
\label{eq_ps}
\Theta(\mbox{\boldmath$n$})=\frac{1}{(2\pi)^2}\int d^2\mbox{\boldmath$l$}\Theta_\textbf{\scriptsize \mbox{\boldmath$l$}}e^{i\mbox{\boldmath$n$}\cdot\mbox{\boldmath$l$}},
\end{equation}
where \emph{\textbf{l}} is a two dimensional wave vector. The PS is then defined as $\langle|\Theta_\textbf{\scriptsize \mbox{\boldmath$l$}}|^2\rangle=C_\textbf{\scriptsize \mbox{\boldmath$l$}}$.
	
We tested that the lack of periodicity does not introduce any error in the computation of the PS by comparing it with what we obtain by filtering the maps with Tukey windows.

\section{Testing the model}
\label{test_model}

In this section we compare the prediction from our model with other simulations and observations, in order to test the robustness of the model. First we perform the comparison for the X-rays and then we focus on the SZ effect.

\subsection{X-ray emission}
\label{Xray_test}

In the past we already studied the X-ray emission properties as predicted from a set of models derived from this same simulation \citep{Ursino10,Ursino11}, and here we report the most significant results. We compare the X-ray estimates from the simulations with observational data. 

At present there is only one direct measurement of the X-ray emission from a WHIM filament by \citet{Werner08} but it has a temperature of the order of $10^7$~K and cannot be accounted for as a representative sample. At the same time, \citet{Galeazzi09} were able to quantify the average X-ray emission from the WHIM in the $0.4-0.6$~keV band as $12\pm4\%$ of the total DXB emission, using a statistical approach. This is consistent with upper limits to the WHIM emission set by the X-ray Quantum Calorimeter (XQC) sounding rocket program \citep{McCammon02}, ROSAT \citep{Kuntz01}, and Chandra \citep{Hickox07}.

\citet{McCammon02} analyze high resolution data of the DXB and set an upper limit to the extragalactic contribution in the energy range $0.380-0.950$~keV of $\sim9$\phot~keV$^{-1}$ ($\sim0.09$\phot \ if we assume an energy interval $\Delta$E$=10$~eV). Our simulation predicts a WHIM surface brightness of $0.059\pm0.004$\phot, within the limits of \emph{XQC}.

\citet{Kuntz01}, as well, set an upper limit to the WHIM emission using \emph{ROSAT PSPC} observations of the north Galactic polar cap. The absorbed extragalactic component in the $380-950$~eV, in fact, has an almost constant value of $\sim10$~keV~cm$^{-2}$~s$^{-1}$~sr$^{-1}$~keV$^{-1}$, corresponding to $\sim8.8$\phot.

As already mentioned, \citet{Galeazzi09} used the AcF to estimate the WHIM contribution of the DXB in the $0.4-0.6$~keV band as $12\pm4\%$. Our model predicts that the WHIM contribution is $17\pm1\%$.

Analyzing the \emph{Chandra} deep fields in the $0.65-1$~keV band, \citet{Hickox07} quantified the unresolved XRB as $(1.0\pm0.2)\times10^{-12}$~ergs~cm$^{-2}$~s$^{-1}$~deg$^{-2}$. In the same band, the WHIM in our simulation has a reassuringly smaller contribution of $(3.6\pm0.3)\times10^{-13}$~ergs~cm$^{-2}$~s$^{-1}$~deg$^{-2}$. 

\subsection{SZ effect}
\label{SZ_test}

\begin{figure}
\plotone{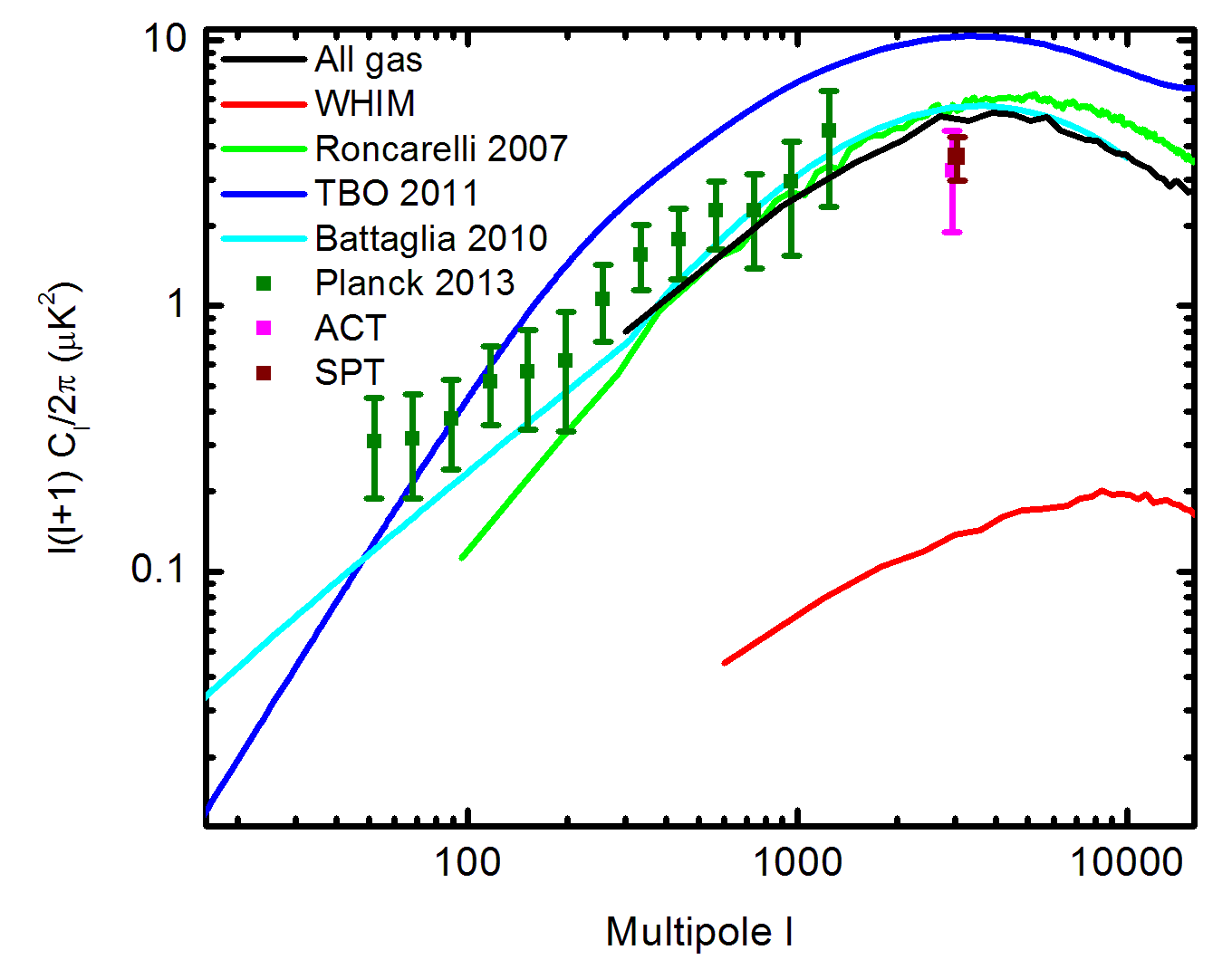}
\caption{The \PS\ of the tSZ signal at 150~GHz from our model of the whole gas (\emph{black}) and the WHIM (\emph{red}). For comparison, the \emph{dot green} line shows the \citet{Roncarelli07} model, the \emph{light blue} line shows the \citet{Battaglia10} model, and the \emph{blue} line shows the \citet{Trac11} model. The estimates from \Planck\ (\emph{dark green}), \ACT\ (\emph{purple}), and \SPT\ (\emph{dark red}) are reported with $1-\sigma$ error bars. At 150 GHz and $l=3000$, the SZ power for \ACT\ and \SPT\ is $3.23\pm1.33$~$\mu K^2$ \citep{Sievers13} and $3.65\pm0.69$~$\mu K^2$ \citep{Reichardt12}, respectively, showing excellent agreement with models. The predicted WHIM signal is very well below the observational constraints for the whole gas. Color version on-line only.
\label{PS_SZ_all_WHIM}}
\end{figure}

In order to test the predictions for the SZ, we compare the WHIM and all-gas \PS\ with the results of other works and with the values measured by \SPT\ \citep{Reichardt12}, \ACT\ \citep{Sievers13}, and \Planck\ \citep{Ade13b}. In figure~\ref{PS_SZ_all_WHIM} the \PS\ from our simulation and from other works.

First we comment the comparison with the work by \citet{Roncarelli07}. The authors, in fact, use our same hydrodynamical simulations \citep{Borgani04}, although they use a completely independent code to generate the SZ maps and to compute the \PS. The excellent agreement for the whole IGM is therefore to be expected if the two procedures are consistent with each other. The difference in the high multipole tail is expected since our work is limited to $z=3$ while they compute the PS up to $z=6$, where there is still a small contribution. \citet{Roncarelli07} also attempted an estimate of the SZ \PS\ from the WHIM but, since they used a different definition from ours (they include the gas with overdensity greater than 1000 that we associate to groups of galaxies), we cannot make a direct comparison.

The model developed by \citet{Battaglia10}, as well, although based on a different set of simulations, has a magnitude close to our model. The \citet{Trac11} standard model has a signal that is about two times higher than what we find but still in our order of magnitude. \citet{Trac11}, however, studied also models that are in the same range of our \PS.

In figure~\ref{PS_SZ_all_WHIM} we also show the values of SZ measured with \ACT\ \citep{Dunkley11, Fowler10, Sievers13}, \SPT\ \citep{Lueker10, Reichardt12}, and \Planck\ \citep{Ade13b}. \ACT\ finds that the SZ power is $\sim 3.23\pm1.33$~$\mu$K$^2$ at $l=3000$ after rescaling their 3-years values to a frequency of 150~GHz (rescaled). \SPT\ 3-years measurements point at a power of $3.65\pm0.69$~$\mu$K$^2$ at $l=3000$ and 150~GHZ. The data reported in the figure have $1-\sigma$ error bars and show an excellent agreement between our model and observations. It is evident, however, that present time instruments do not have the required sensitivity to detect the WHIM signal, since it is about a factor 10 below the current detections.

Recently, \citet{Suarez13} developed a model based on the lognormal distribution of the baryon density, \citet{Genova13} compared that model with \emph{WMAP} (7 years) and \SPT. The two works agree with each other, but strongly disagree with our results. They find that the \PS\ of the WHIM SZ is at $l\approx300-500$ and has an amplitude much higher than the scale we chose for figure~\ref{PS_SZ_all_WHIM}. The very different results could be explained if we consider the differences between our model and the model of \citet{Suarez13}. The log-normal formalism adopted by \citet{Suarez13} does not consider the effects of metals and how metal cooling affects the temperature and density evolution of the WHIM structures but rather generates average properties of the WHIM. \comRef{Temperature, as well, is defined as a function of density instead of an independent variable.} Our model, instead, follows the evolution of the baryons and uses the physical properties of gas elements to predict the SZ from the WHIM. The different approach could give different statistical properties of the WHIM structures, and therefore of the SZ \PS.
  
\section{Angular correlations}
\label{large_scale_XSZ}

\subsection{X-rays}
\label{X_corr}

\begin{figure*}
\plottwo{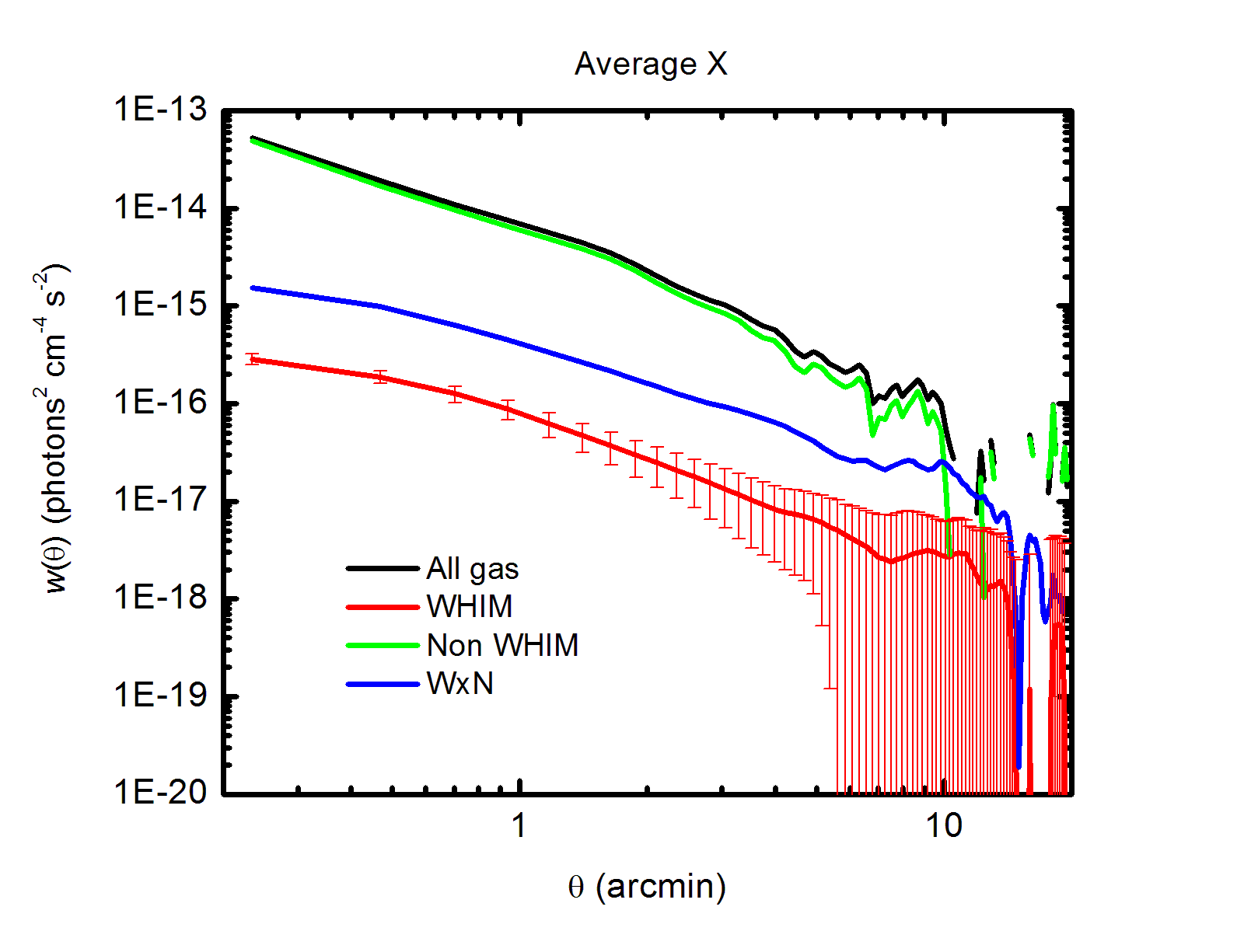}{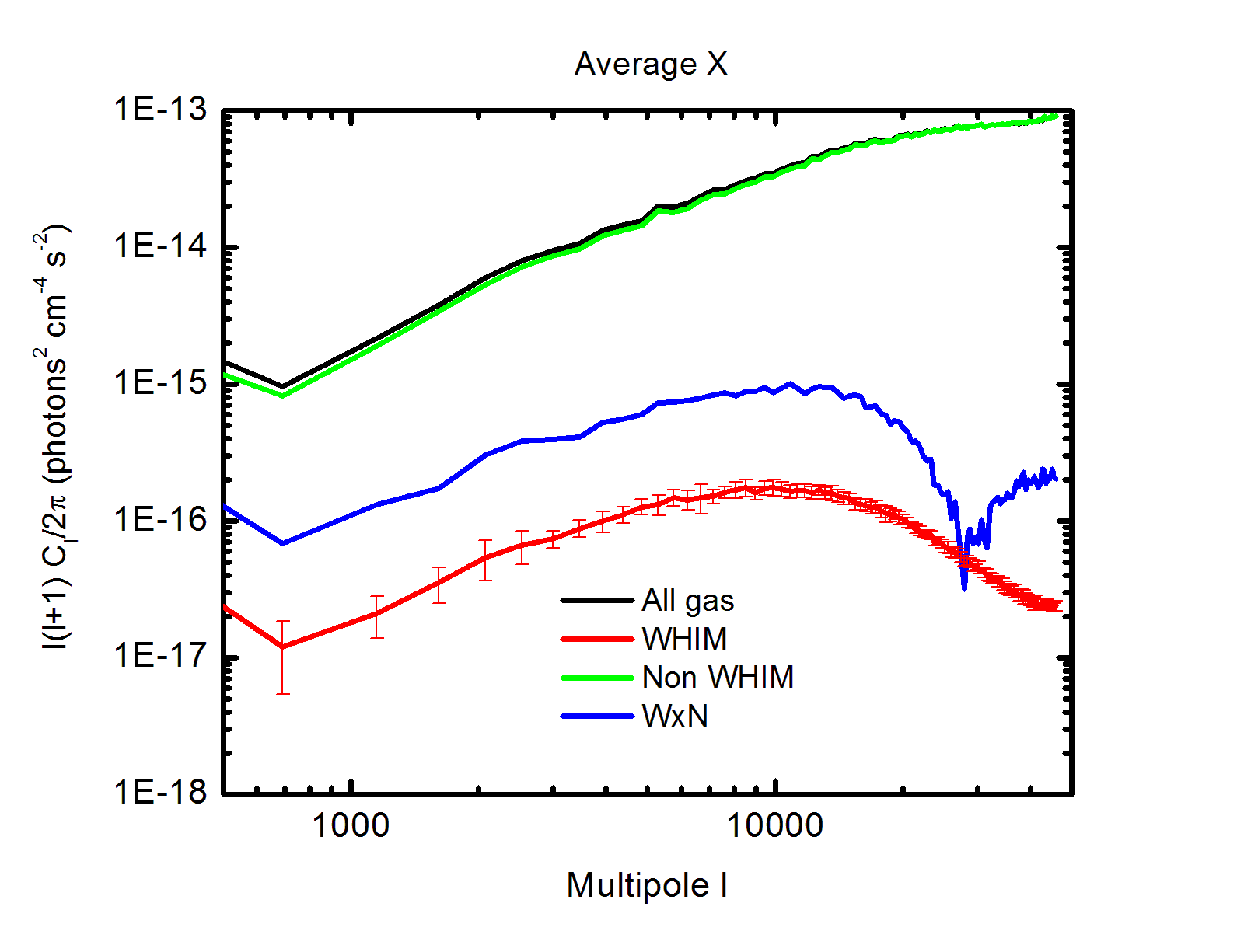}
\caption{AcF (\emph{left}) and \PS\ (\emph{right}) of the X-ray emission for the whole gas (\emph{black}), for the WHIM (\emph{red}), and for the non-WHIM (\emph{green}). We also show the cross-correlation term between WHIM and non-WHIM (\emph{blue}). The figures show that the correlation signal is largely dominated by the non-WHIM component. The WHIM signal, however, is significantly different from zero. Notice that the trough at $l\sim25000$ in the cross-correlation signal is actually an artifact since we set the correlation positive everywhere:  small-scale (high-l) values are negative. For sake of clarity we show the variance only for the WHIM signal, the variance of the  other signals, however, is comparable with that of the WHIM. Color version on-line only.
\label{PS_XXav}}
\end{figure*}

In figure~\ref{PS_XXav} we show both the AcF and the \PS\  of the X-rays emitted by all of the gas, WHIM, and non-WHIM. We also show the correlation term between the WHIM and the non-WHIM, since this is a component of the total correlation signal as well. As expected, the non-WHIM component makes for almost all the signal. The cross-correlation term, as well, is much stronger than the WHIM contribution. We remind the reader that, from an observational point of view, it is possible to remove most of the non-WHIM contribution, leaving the WHIM as the dominant term. In \citet{Ursino11}, in fact, it is shown that by removing the brightest pixels from an ``all-gas'' map, regardless of their association to known objects, the WHIM becomes the dominant contributor; in particular removing the 50\% brightest pixels the WHIM will account for $\sim75\%$ of the total X-ray signal. In real observations, multi-wavelength observations can be used to remove virilized structures from the observation, with even better rejection of non-WHIM gas (e.g., M. Galeazzi et al. 2014, in preparation). We note that, for this work, we tested the same methodology for SZ maps and obtained remarkably similar results: the signal of the WHIM makes up for $\sim75\%$ of the total AcF after removing 50\% of the pixels with the highest temperature fluctuations. 

From the AcF plot, we see that the total signal falls as $\sim \theta^{-2}$, while the WHIM signal is shallower. This means that in general the all-gas X-ray signal is collected in much smaller and brighter structures than the WHIM signal.  

The \PS\ plot shows that the WHIM signal has a maximum at $l\sim10000$. The signal from the IGM, instead, is peaked at a higher value that is outside the multipole range that we have set (or equivalently, at an angle smaller than the resolution of our simulation). The correlation term has a peak at $l\sim15000$ and then it quickly drops to negative values (our algorithm computes the absolute values of \PS, we verified that the dip at $l\sim30000$ is indeed the turning point, after which the signal from the correlation term is actually negative). 

We note that high-$l$ power spectrum for the WHIM / non-WHIM cross-correlation is negative, while the small scale AcF shows positive correlation. This possibility may be understood by the physical, geometrical association of the non-WHIM signal (at cluster halo cores) and the WHIM signal (in the halo outskirts). Projecting a halo from three-dimensions to two, the WHIM emission concentrates around the halo center, but is somewhat depressed along the line of sight through the excised, non-WHIM halo core. Since both the WHIM and non-WHIM gas are associated with the halo, on large scales, the two are correlated, as seen in the low-$l$ power spectrum. If large scales are suppressed, the depressed WHIM emission near the halo center anti-correlates with the non-WHIM emission at the same location. This explains the negative cross-power spectrum at high-$l$ in figure~\ref{PS_XXav}: in this sense, the power spectrum is showing the variation from the local average. However, in the measurement of the AcF, larger angular scales are retained. At the halo center, both the WHIM and non-WHIM signal are above their respective global means, and thus they have positive correlation.

\subsection{SZ}
\label{SZ_corr}

\begin{figure*}
\plottwo{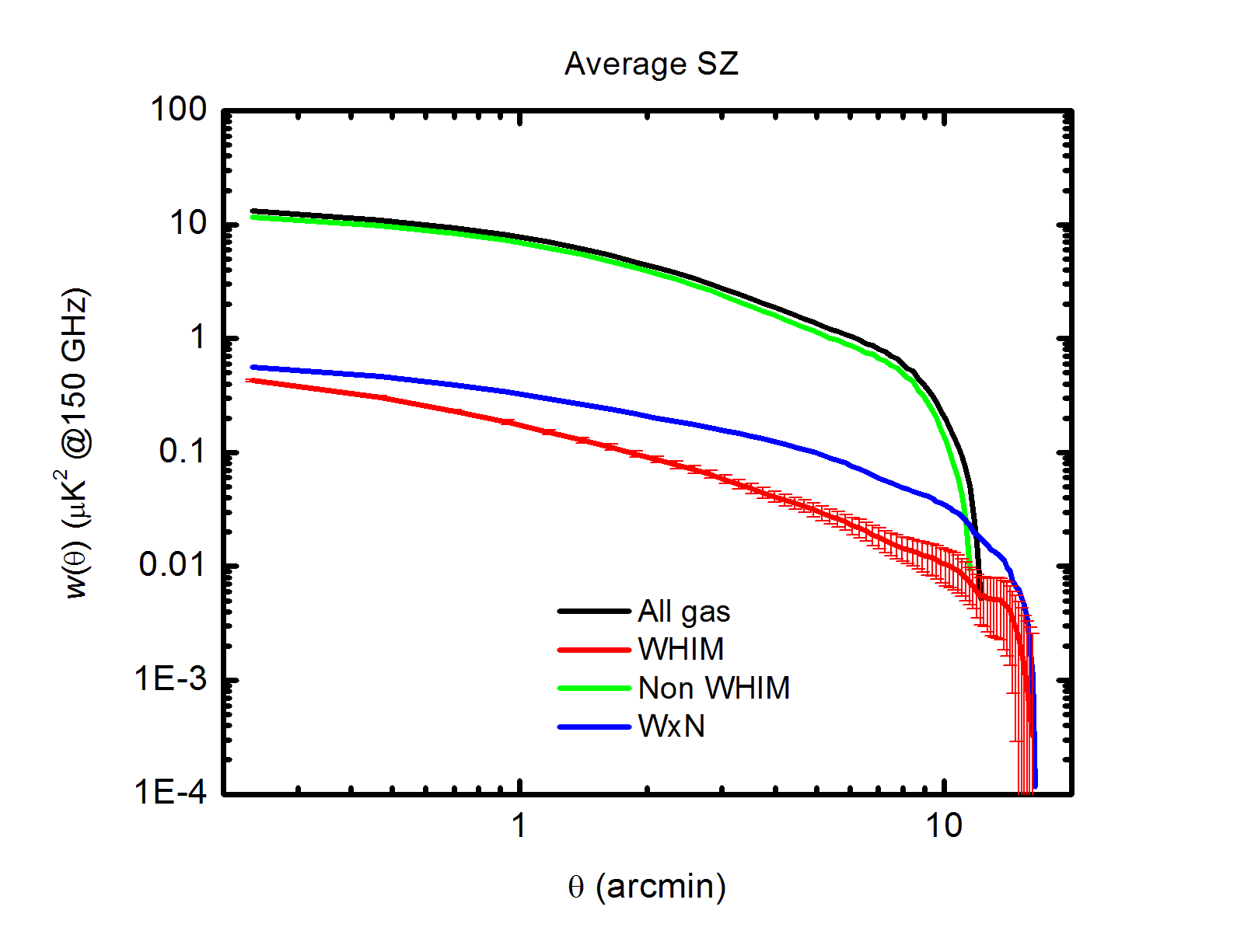}{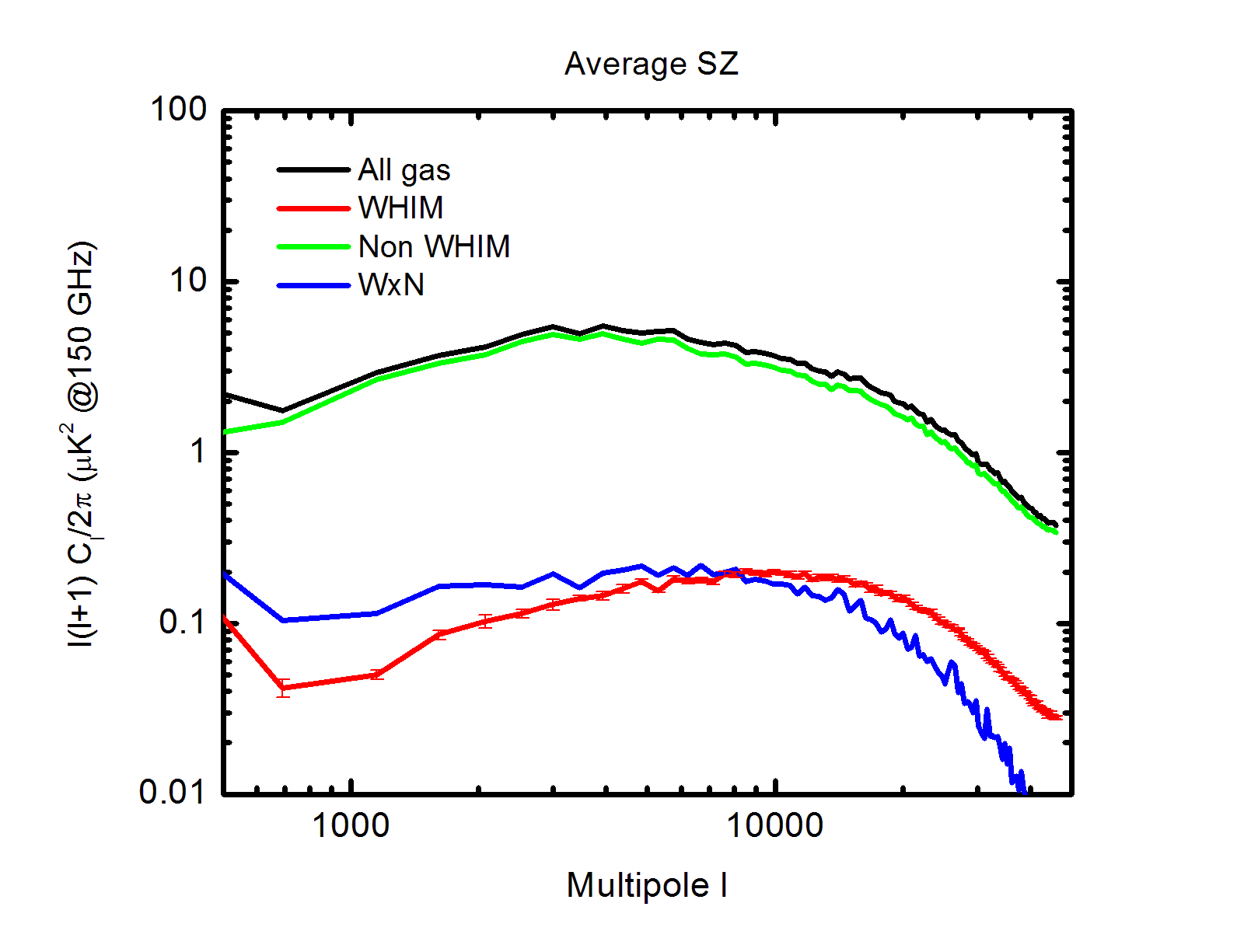}
\caption{AcF (\emph{left}) and \PS\ (\emph{right}) of the SZ effect at 150~GHz for the whole gas (\emph{black}), for the WHIM (\emph{red}), and for the non-WHIM (\emph{green}). We also show the cross-correlation term of WHIM-non WHIM gas (\emph{blue}). The figures show that the correlation signal is largely dominated by the non-WHIM component. The WHIM signal, however, is significantly different from zero. For sake of clarity we show the variance only for the WHIM signal, the variance of the  other signals, however, is comparable with that of the WHIM. Color version on-line only.
\label{PS_SSav}}
\end{figure*}

Figure~\ref{PS_SSav} is the analog of figure~\ref{PS_XXav}, this time we show the angular properties of the WHIM, non-WHIM, and all-gas SZ and the correlation between the WHIM and non-WHIM. Similarly to the X-ray emission, the SZ signal is dominated by the non-WHIM component. The angular properties, however, are rather different from those of the X-ray maps.

The SZ AcF signal of both WHIM and IGM is shallower than the X-ray signal up to $\theta\sim10'$ and then it quickly drops to negative values. This means that the same structure appears smaller in X-rays than in the SZ. The SZ contrast between the central part of a structure and its outskirt, as well, is much smaller than in the case of X-rays. We also note that the cross-correlation between WHIM and non-WHIM behaves differently. In X-rays it is about an order of magnitude stronger than the WHIM AcF at all angles in the range of interest. The SZ cross-correlation, instead, is almost identical to the WHIM at small angles ($\theta<1'$) and then it becomes $\sim3$ times larger than the WHIM AcF at angles of the order of a few arcminutes. This fact can be explained if we consider that the surface covered by an object is larger in the SZ maps than in X-rays. Where WHIM and non-WHIM overlap (small angles) their signals have almost the same amplitude and the cross-correlation is similar to the WHIM AcF. At separations of few arcminutes, instead, there is real distinction between the two phases, giving rise to a cross-correlation stronger than the WHIM AcF.

Similarly, also the trends of the \PS\ plot show differences between the X-ray emission and the SZ. The peak for the all-gas signal is $l\sim5000$. The WHIM peak, instead, is at $l\sim10000$ (like in X-rays), but the shape of the curves differ from X-rays (in X-rays there is higher difference between the values at the peak and at the lowest multipoles). Since smaller multipoles correspond to large angles, we see the same picture described by the AcF: in the SZ we see that the structures are larger and smoother.

In order to understand the different angular properties of the IGM as seen in X-rays and through the SZ effect, we have to consider the physical mechanisms that rule the radiation in the two different energy bands. There are two variables that explain why the angular distribution of the X-ray emission differs from that of the SZ effect. One is distance: SZ is not affected by distance, while X-rays go as the inverse squared of distance (although in both cases the angular diameter is affected by distance and the signal sources are different at earlier times). This means that, while we cannot see the distant X-ray signal from the WHIM, the SZ contribution is still present, characterizing the small angle distribution of distant objects. The other parameter is density: the X-ray emission depends on the square of density, while the SZ effect is roughly linear. Usually the densest regions (stronger X-ray emitter) have a limited size (typically enveloping virialized structures). As a result the X-ray emission enhances more the contribution from relatively smaller regions. Since the X-ray AcF peak is shifted to smaller angles, it appears evident that the density dependence is a stronger factor than 
the distance. 

\subsection{X-rays/SZ cross-correlation}
\label{XSZ_corr}

\begin{figure*}
\plottwo{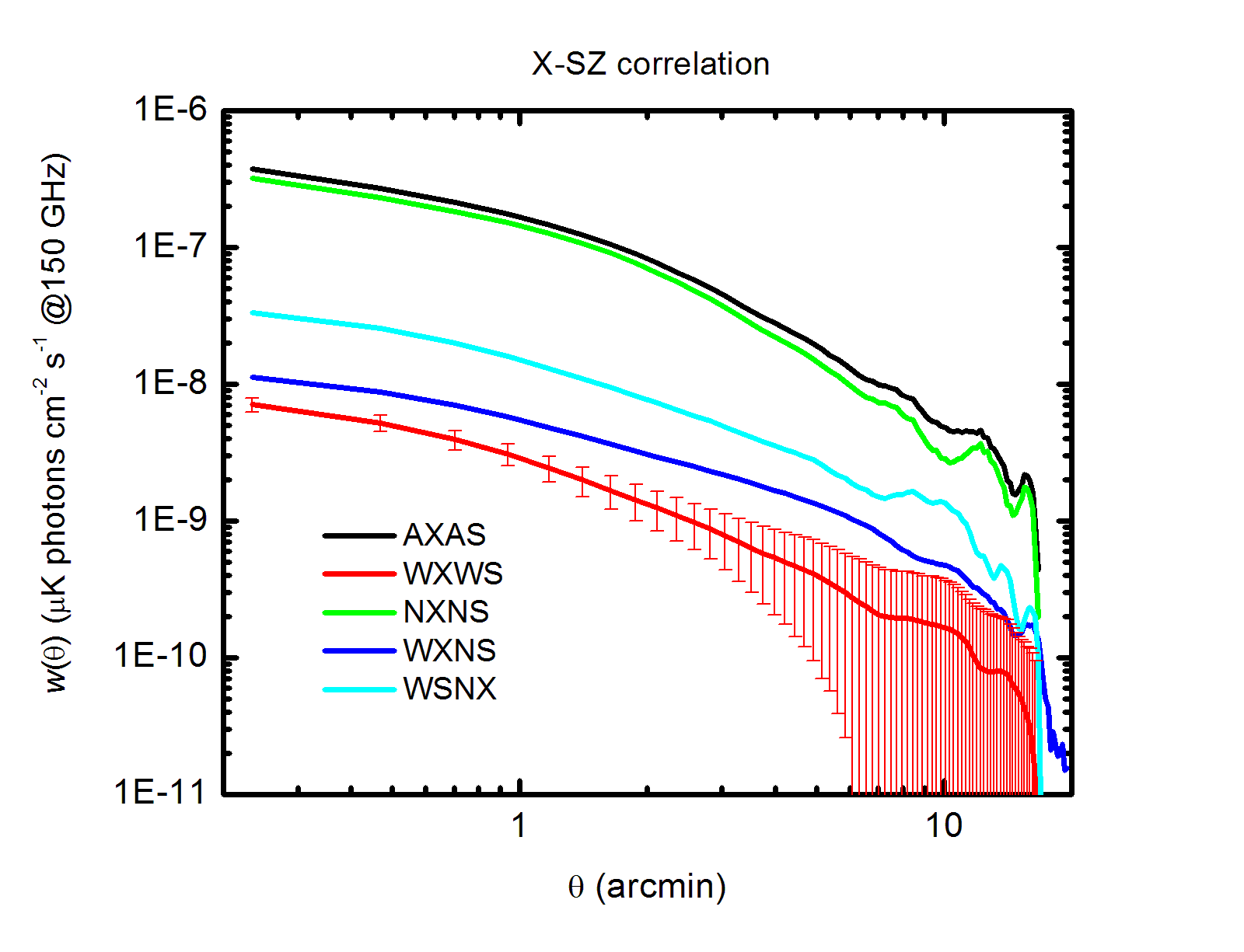}{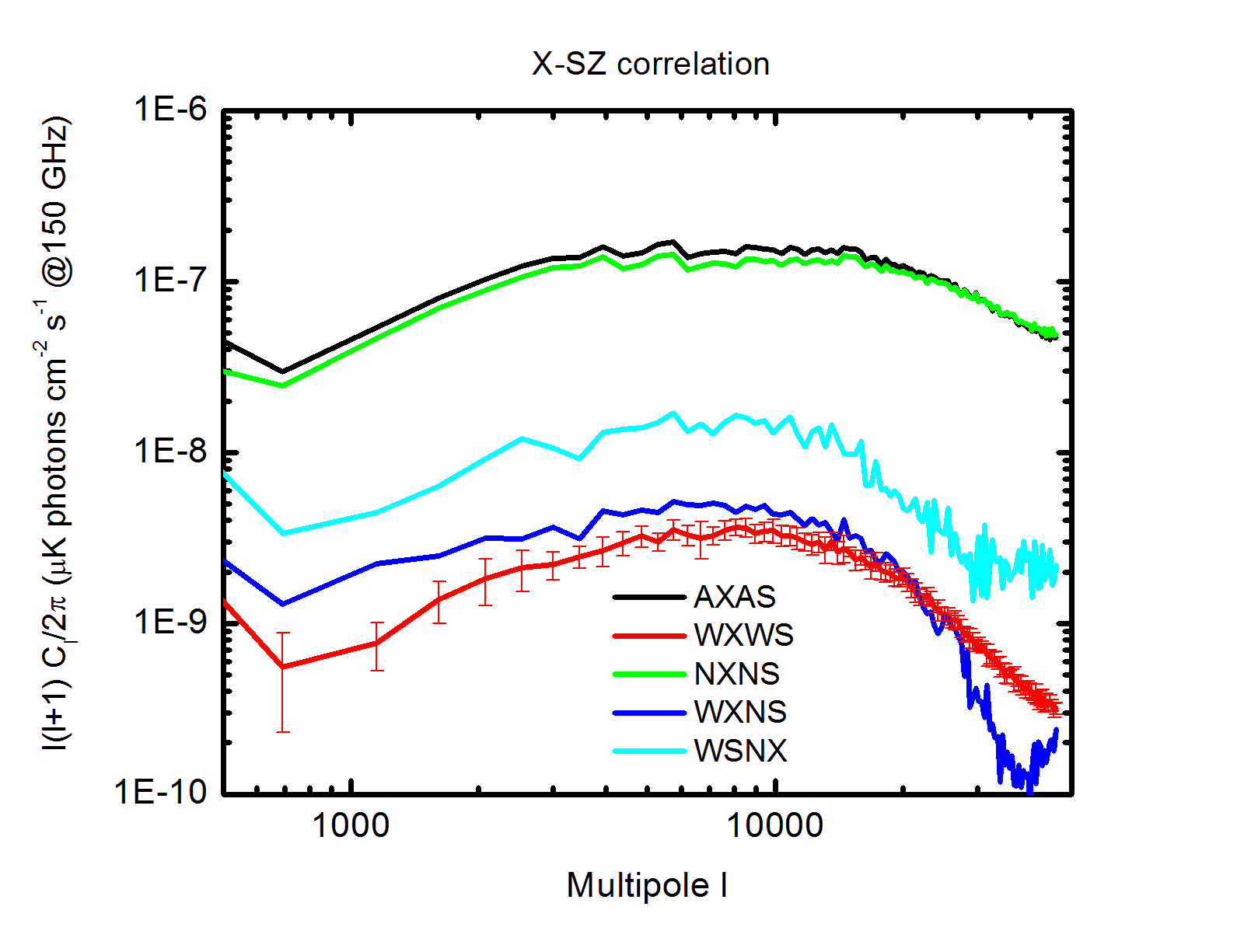}
\caption{SZ (at 150~GHz) X-ray correlated AcF (\emph{left}) and \PS\ (\emph{right}) for the whole gas (\emph{black}, label \emph{AXAS}), for the WHIM (\emph{red}, label \emph{WXWS}), and for the non-WHIM (\emph{green}, label \emph{NXNS}). There are also the cross-correlation terms of WHIM X-ray vs. non-WHIM SZ (\emph{blue}, label \emph{WXNS}) and WHIM SZ vs. non-WHIM X-ray (\emph{light blue}, label \emph{WSNX}). The figures show that the correlation signal is largely dominated by the non-WHIM component. The WHIM signal, however, is significantly different from zero. The correlation signal between WHIM and non-WHIM is stronger when the WHIM is observed in X-rays. For sake of clarity we show the variance only for the correlation of the X-ray and SZ WHIM signal, the variance of the  other signals, however, is comparable with that of the WHIM. Similarly to figure~\ref{PS_XXav}, the trough at $l\sim35000$ in the \emph{WXNS} signal is an artifact and small-scale (high-l) values are negative. Color version on-line only.
\label{PS_XSav}}
\end{figure*}

Finally, we investigate if there is correlation between X-ray and SZ maps. The visual inspection of the two maps in figure~\ref{maps} shows that there are similarities. However, the need of this investigation comes from the more practical considerations of finding new independent ways of identifying the elusive signal of the WHIM. As we stated in section~\ref{introduction}, there are very few detections of the WHIM in absorption, and all of them, at some level, are questionable. In emission there is only the detection of a filament \citep{Werner08}, although it is so hot and dense that is at the limits of the definition of WHIM, and the detection of the AcF signal in \emph{XMM-Newton} fields \citep{Galeazzi09}. A statistical approach of the angular properties of the WHIM through the analysis of SZ maps is still beyond feasibility, as shown in figure~\ref{PS_SZ_all_WHIM}. However, we want to verify if, by combining X-ray and SZ maps, it is possible to identify some characteristic features of the WHIM.

In figure~\ref{PS_XSav} we show the cross-correlation between X-ray and SZ maps. First of all, we want to stress that the X-ray and SZ signals are actually anti-correlated (stronger X-ray emission correlates with larger negative temperature fluctuations) and that we plot the absolute value of the cross-correlation so that it is easier to directly compare it with the X-ray and the SZ autocorrelations plots. However we also note that the sign of the temperature fluctuations (and therefore the flip between correlation and anti-correlation) is frequency dependent: for $\nu<217$~GHz, in fact, temperature fluctuations have negative sign (we adopted 150~GHz) and become positive for $\nu>217$~GHz. Adopting a more general description of SZ in terms of the $y$ parameter instead of $\Delta T/T$ we would obtain direct correlation between X-rays and SZ maps. Besides the ``all-all'' and WHIM-WHIM terms, we also plot the cross-terms with non-WHIM, to verify which one is the dominant factor. As expected, the non-WHIM component is the dominant one, and the WHIM term is the smallest. Looking at the cross-terms between WHIM and non-WHIM, it is interesting to note that the correlation between the WHIM SZ and the non-WHIM X-rays is almost an order of magnitude stronger than the correlation between WHIM X-rays and non-WHIM SZ. Furthermore, the two curves have different shapes, with the ``WHIM SZ non-WHIM X-ray'' term showing stronger correlation at small angles. The different amplitude comes from the different ratio between the non-WHIM and WHIM signal in X-rays and SZ, with the first being about two orders of magnitude and the second a little more than one order of magnitude (see figures~\ref{PS_XXav} and \ref{PS_SSav}). While the cause for the different shapes can be found in the fact that the non-WHIM X-ray is much more concentrated in the halo than the non-WHIM SZ, while for the WHIM the difference is not so remarkable. 

Both the WHIM and the all-gas signal peak at $l\sim10000$ and the cross-correlation peaks accordingly. While the WHIM does not change significantly the peak position compared to X-ray auto-\PS, the peak for the IGM is in a very different position, giving a way to disentangle the WHIM and non-WHIM signal by comparing the X-ray auto-correlation with the cross-correlation between X-ray and SZ maps. In the conclusions we discuss the removal of the non-WHIM signal by removing point sources and clusters.

\section{Redshift Evolution}
\label{Redshift_Evolution}

In this section we investigate how X-rays and the SZ evolve with the redshift. In order to understand where the signal comes from, first we analyzed the redshift evolution of gas phases.

\begin{figure}
\plotone{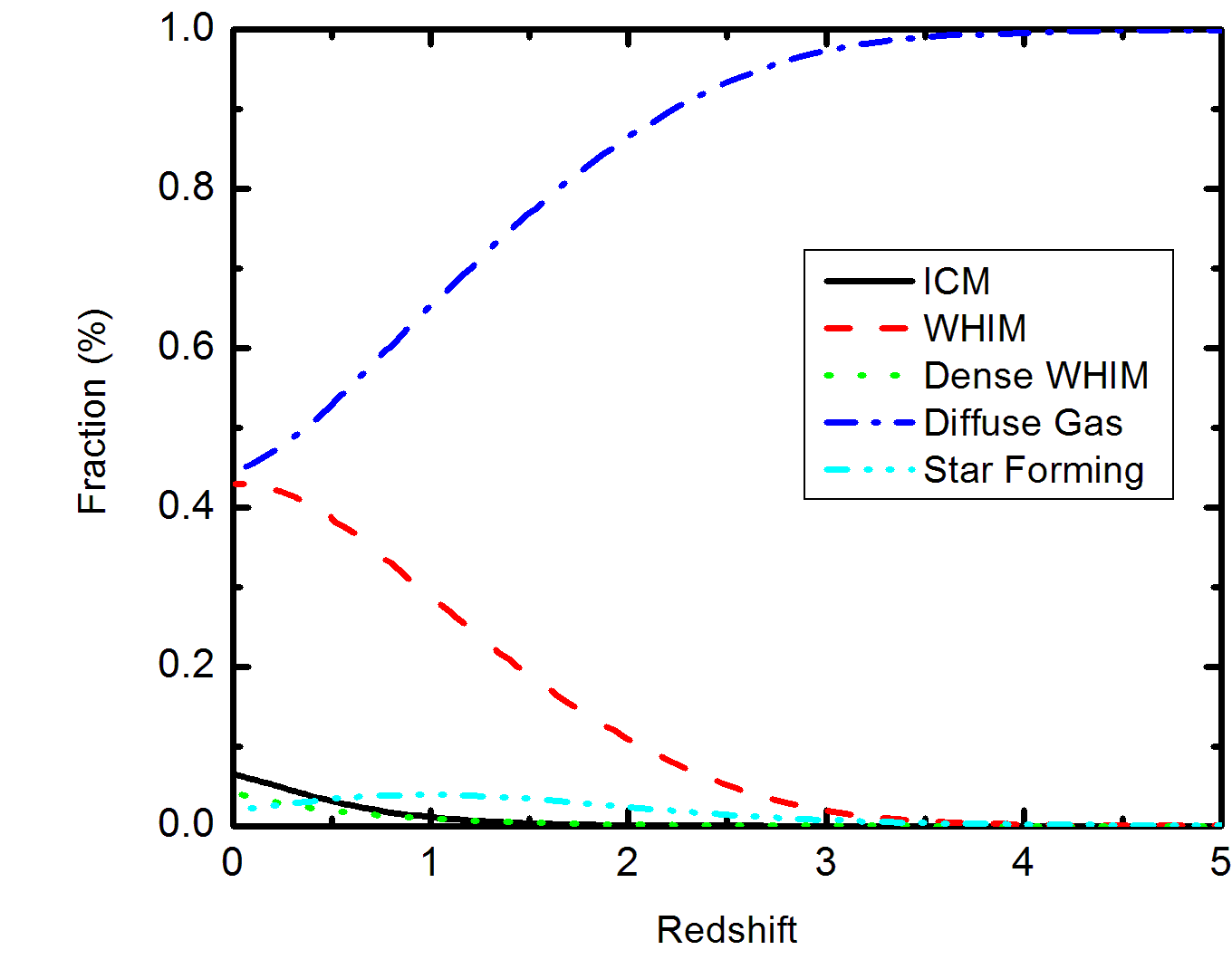}
\caption{Evolution of gas phases: hot gas (\emph{black}), dense WHIM (\emph{green}), WHIM (\emph{red}), diffuse gas (\emph{blue}), and star forming gas (\emph{light blue})). At earlier time the baryonic matter is almost only in the diffuse phase. At present time the WHIM account for $\sim40\%$ of the baryons, like the diffuse gas. The other phases account for a much smaller fraction. Color version on-line only.
\label{phase_evolution}}
\end{figure}

We divided the gas particles in each snapshot of the \citet{Borgani04} simulation in five different phases, besides WHIM, non-WHIM and Intercluster Medium (ICM, $T>10^7$~K), we also defined the diffuse gas ($T<10^5$~K and $\rho<1000\rho_b$) and the star forming phase (defined as $T<10^5$~K and $\rho>1000\rho_b$. Following the redshift evolution from $z=5$ to $z=0$ in figure~\ref{phase_evolution}, we see that initially almost all the gas was in the form of the diffuse gas (\emph{blue}). With time, due to gravity, the gas collapsed in more compact structures, eventually ending up in virialized objects. At present day the simulation predicts that less than $\sim45$\% of the gas is in the form of the diffuse gas, and almost the same amount is WHIM (\emph{red}). Less than 10\% of the gas is in the form of clusters and groups (ICM and dense WHIM). Other hydrodynamical simulations see similar results, and the fractions in the various gas phases differ by a few percent or less. 

\begin{figure}
\plotone{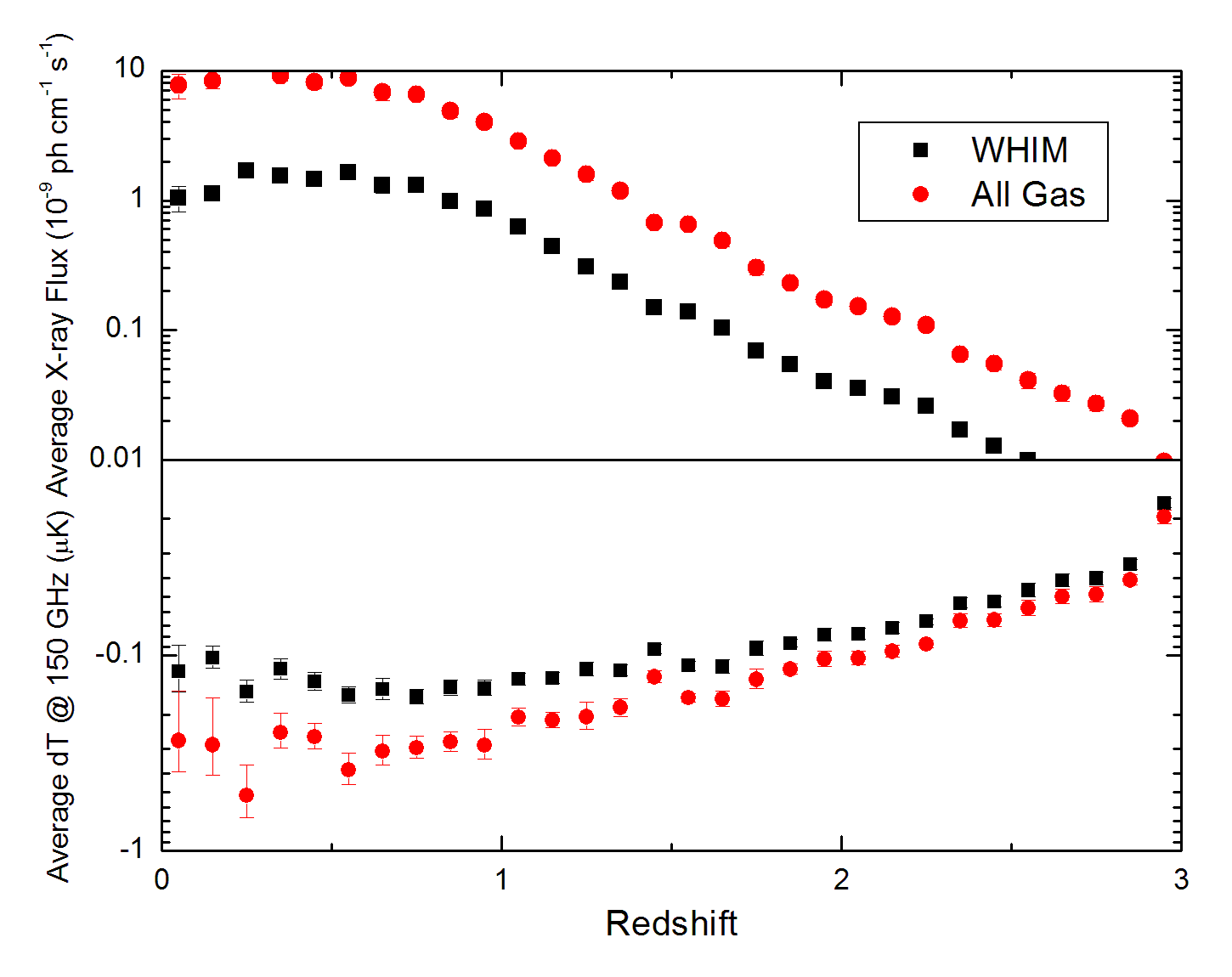}
\caption{X-ray flux and temperature fluctuations for WHIM (\emph{black squares}) and all the gas (\emph{red circles}) as a function of redshift. The X-ray WHIM signal becomes negligible at redshift higher than 1.5. The WHIM SZ signal, instead, is still relatively large almost up to redshift 3. Color version on-line only.
\label{X_SZ_vs_z}}
\end{figure}

Following the analysis of the gas phase evolution, we computed the average X-ray intensity and temperature fluctuations in both WHIM and all-gas maps separated by redshift intervals $\Delta z=0.1$. SZ and X-ray evolution show the same pattern: in both cases the total IGM signal is up to $\sim10$ times stronger than the WHIM signal, in both cases the intensity is stronger at smaller redshift (just like the fraction of WHIM and ICM becomes more and more important), and in both cases most of the growth happens at $z>0.5$. In figure~\ref{X_SZ_vs_z} we can follow the redshift evolution of both X-ray emission (top panel) and average SZ temperature decrement (bottom panel). As we stated before, for $z>2$, we see that the contribution from X-rays is negligible. Temperature fluctuations, instead, are still present but the comparison with the results of \citet{Roncarelli07} reassures us that the contribution from $z>3$ is marginal.	

\begin{figure*}
\plottwo{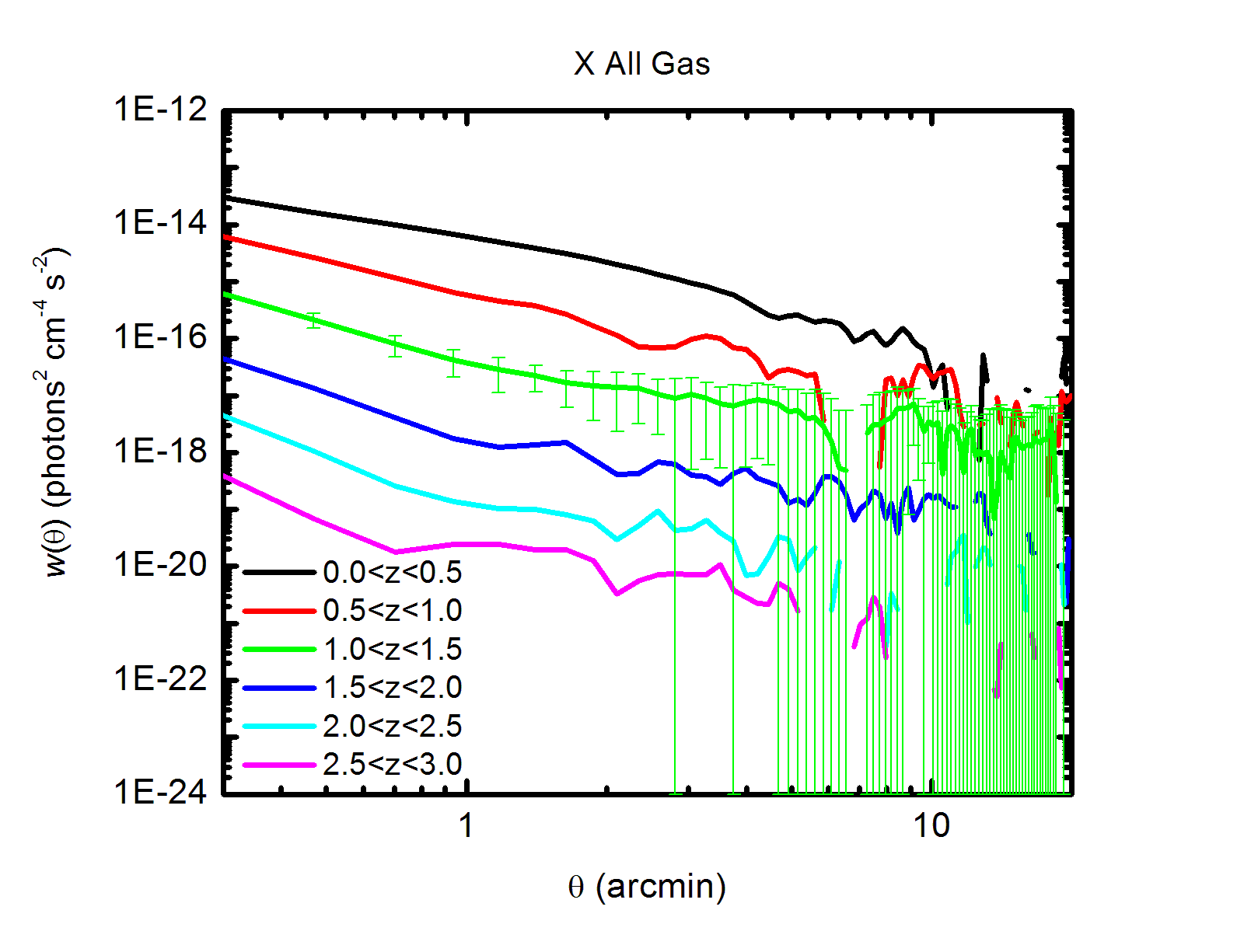}{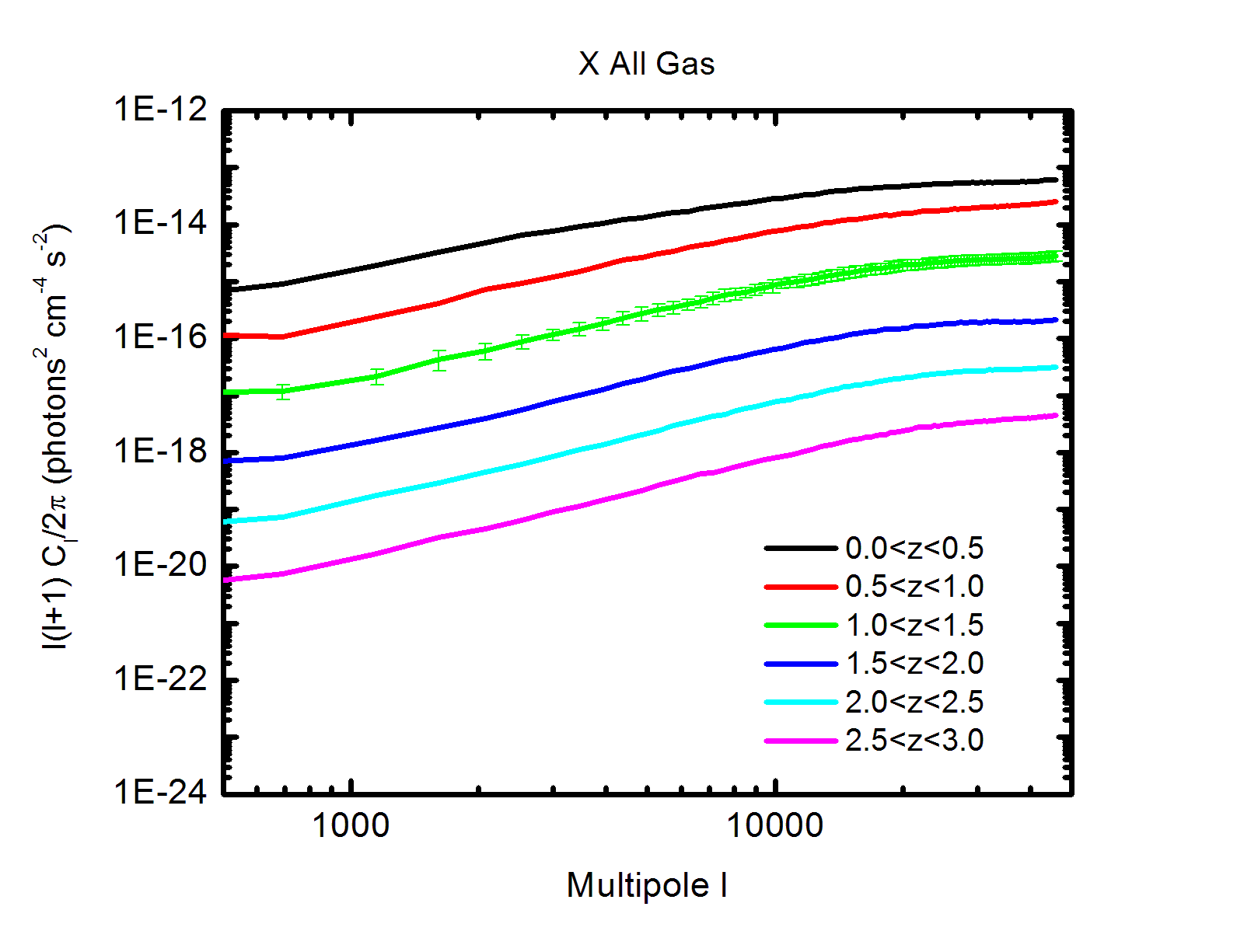}
\plottwo{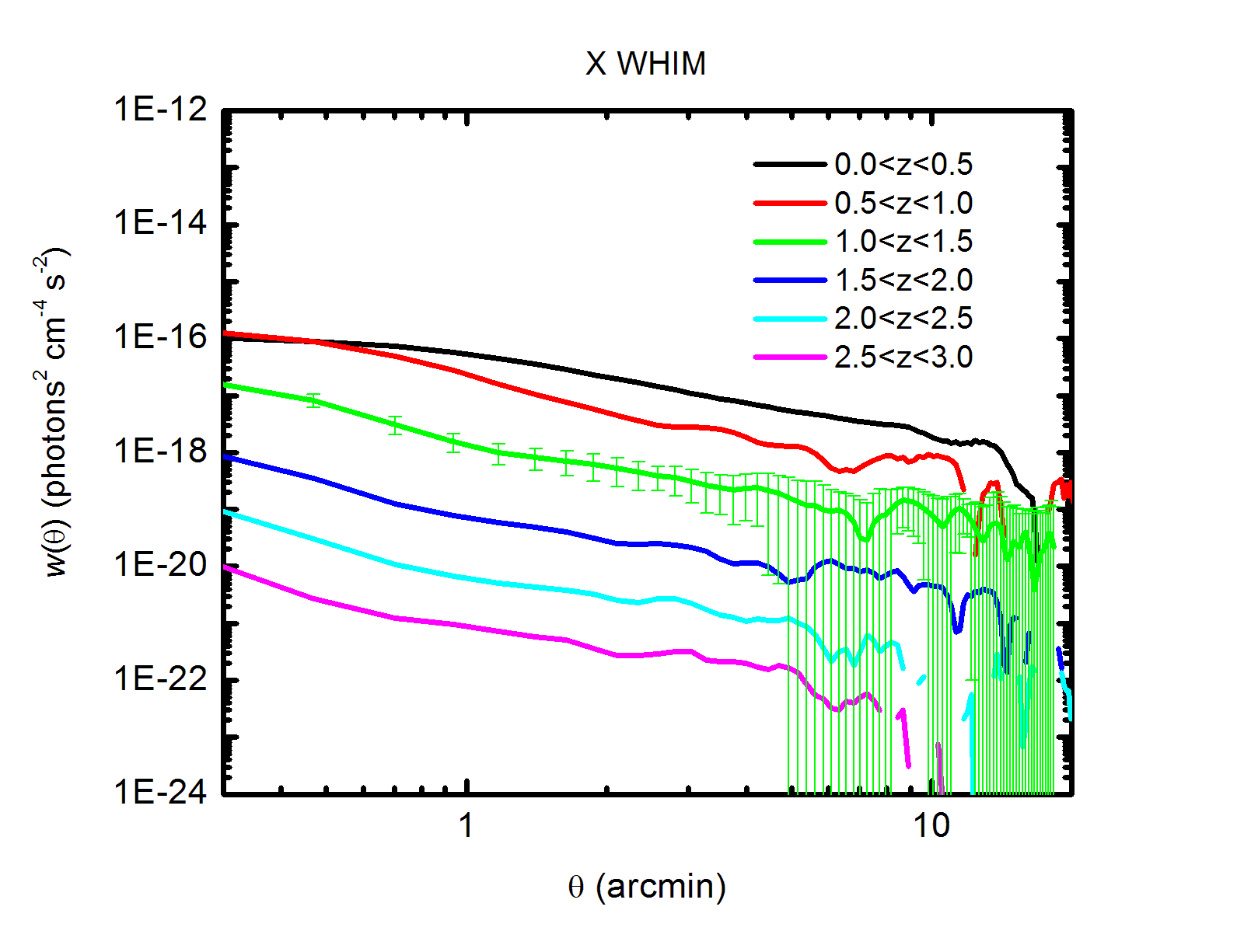}{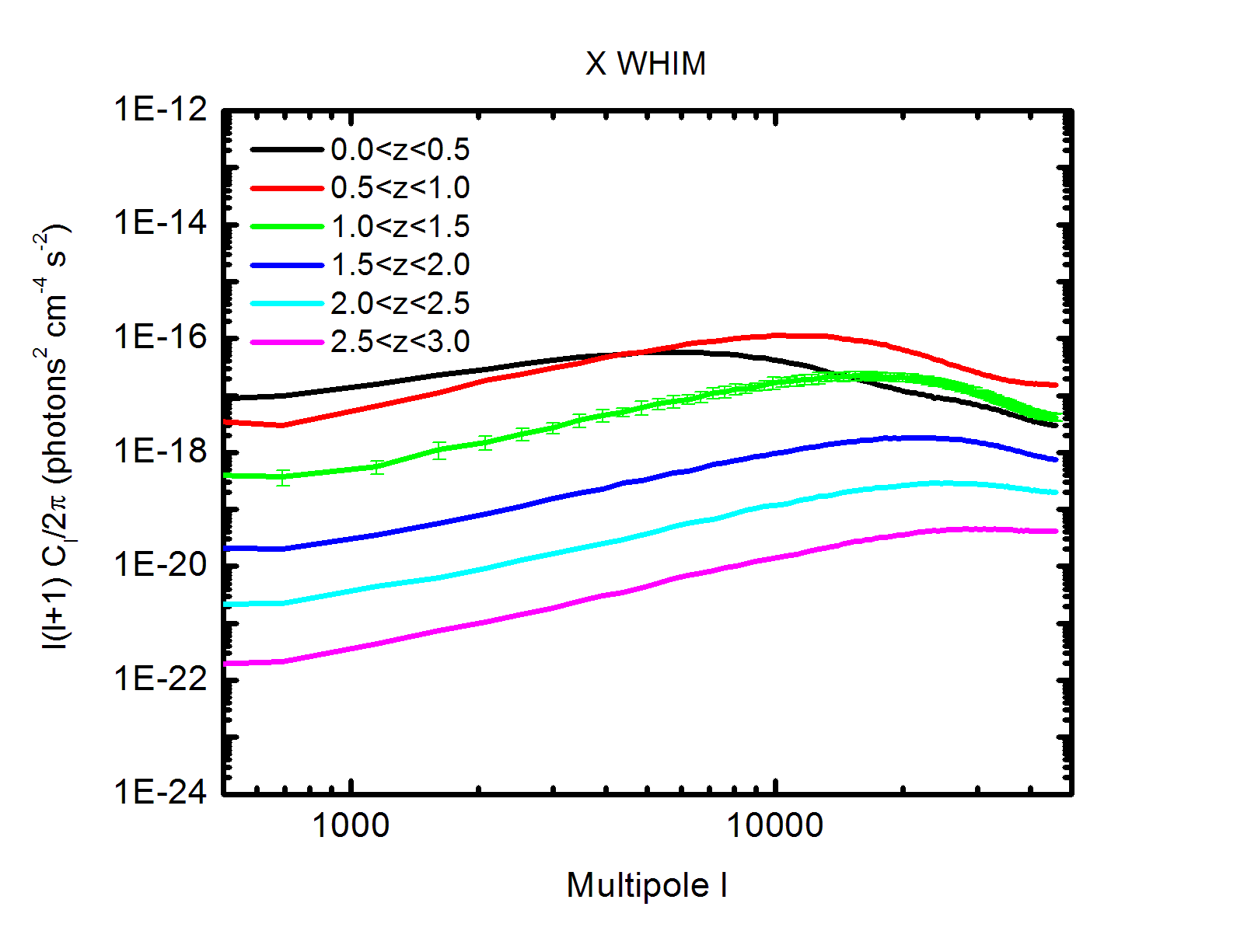}
\caption{Contribution to all-gas (\emph{top}) and WHIM (\emph{bottom}) X-rays emitted at equal redshift intervals. Notice that the WHIM and all-gas plots have the same scale for an easier and direct comparison. The figures show that the correlation signal is larger at small redshift. The WHIM signal, though smaller than the all-gas signal, is significantly different from zero. The maxima of the \PS curves are at lower poles for lower redshifts. For sake of clarity we show the variance only for the [$1.0<z<1.5$] interval, the variance of the other signals, however, is comparable with that of the [$1.0<z<1.5$] interval. Color version on-line only.
\label{A_X_zs}}
\end{figure*}

Since both average X-ray intensity and SZ temperature fluctuations evolve with time, we also investigate how baryons at different distances contribute to the correlation signal, to see if the detected signal comes from evenly distributed regions, or if there is a preferential emission region. In order to do this, we generated maps of the signal coming from baryons within intervals equally spaced in redshift, with $\Delta z=0.5$.

\subsection{Redshift evolution of the X-ray signal}
\label{X_corr_evo}

In figure~\ref{A_X_zs} we show the evolution of the angular properties of the WHIM and all-gas X-ray signals emitted in regions with $\Delta z=0.5$ for $z<3$. At early time ($z\ge1$) the amplitude of the WHIM signal increases quickly at an almost steady rate of about two decades per redshift unit, after which it shows smaller variations (in agreement with the average values shown in figure~\ref{X_SZ_vs_z}). The IGM signal has a similar behavior, although it shows stronger evolution than the WHIM also at late times ($z<1.0$). The shape of the signals changes with redshift, as well. The all-gas PS does not show any peak in our range (the peak is likely located at some $l>50000$), but from the AcF plot we can see that the slope is steeper at higher redshift, pointing to smaller characteristic angles. The \PS\ plot for the WHIM, instead, shows peaks that shift from $l=5000$ (lower redshift) to $l=50000$, suggesting that the structures have smaller angular size when they are further away.

In X-rays, the closer the redshift interval, the more it contributes to the total signal, both in the case of the WHIM and for the general IGM. This is due to distance, to evolution of large scale structures, and to selection of spectral features. X-rays are inversely proportional to the distance squared, therefore the intensity should drop quickly with distance, although this effect should be compensated by the fact that the volume increases with distance. However, as we see in figure~\ref{phase_evolution}, the amount of baryons in virialized or partly collapsed structures, is much less at larger redshifts, with the result that there are less baryons capable of emitting X-rays. By $z<0.5$, growth in the WHIM phase begins to level off, possibly indicating that the WHIM is reaching a steady state equilibrium, where baryons keep accreting into the WHIM but almost as many become so hot and dense to go into the definition of groups and clusters. Furthermore, the characteristic O\elem{VII} and O\elem{VIII} lines (the largest contributors to the WHIM X-ray emission) emitted at distances larger than $z\simeq0.5$ are shifted at energies below 0.4~keV (the lower limit of our energy band). 

\subsection{Redshift evolution of the SZ}
\label{SZ_corr_evo}

\begin{figure*}
\plottwo{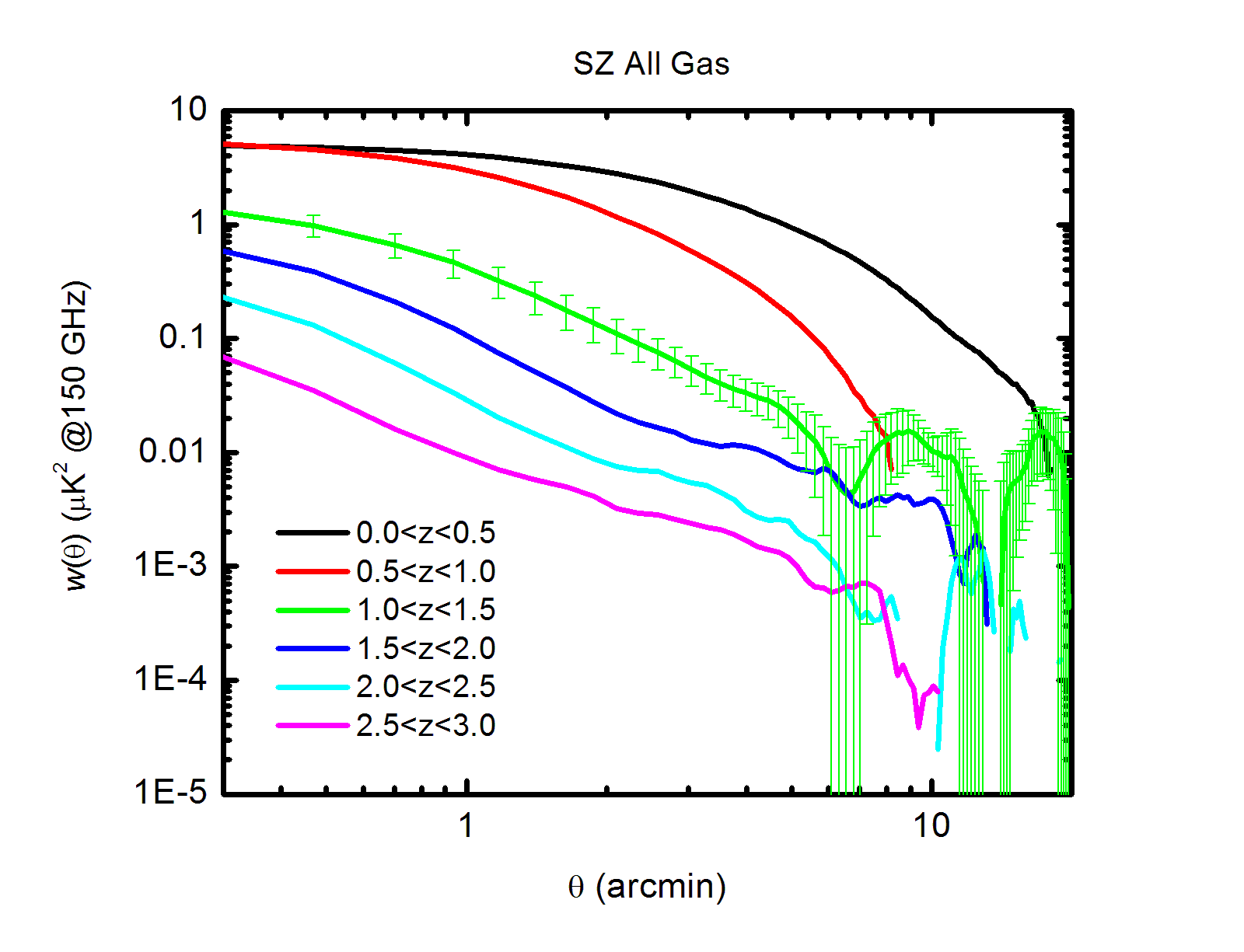}{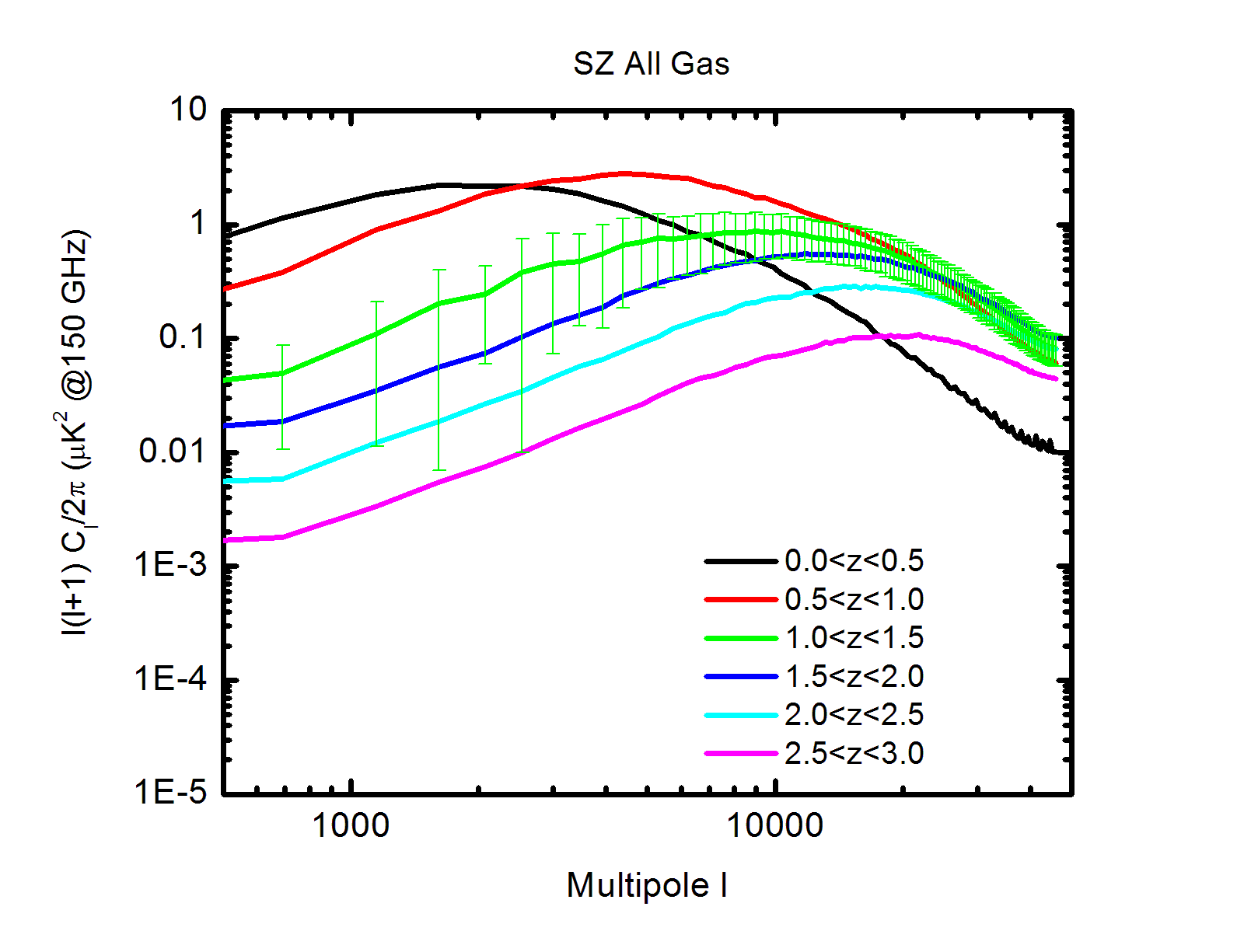}
\plottwo{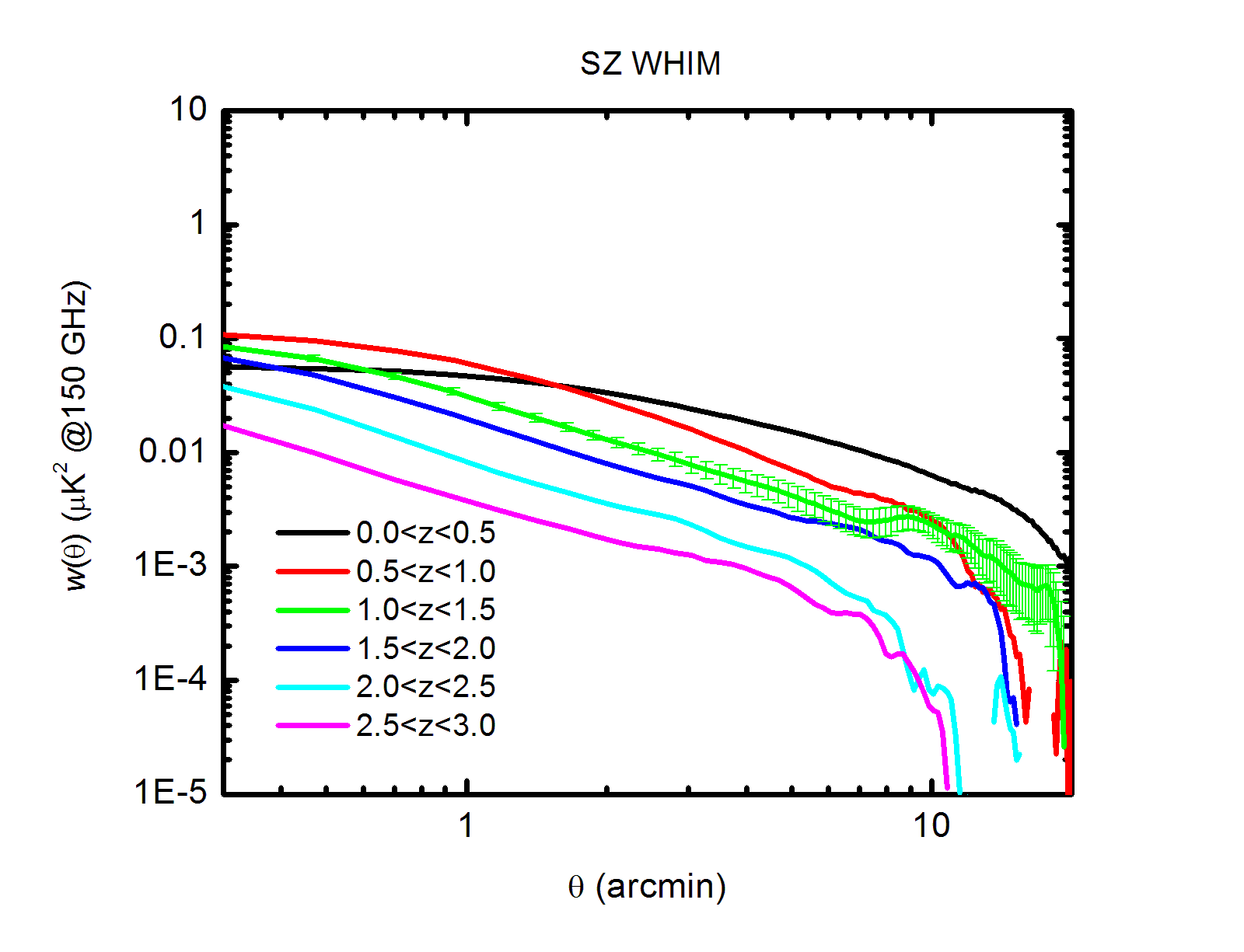}{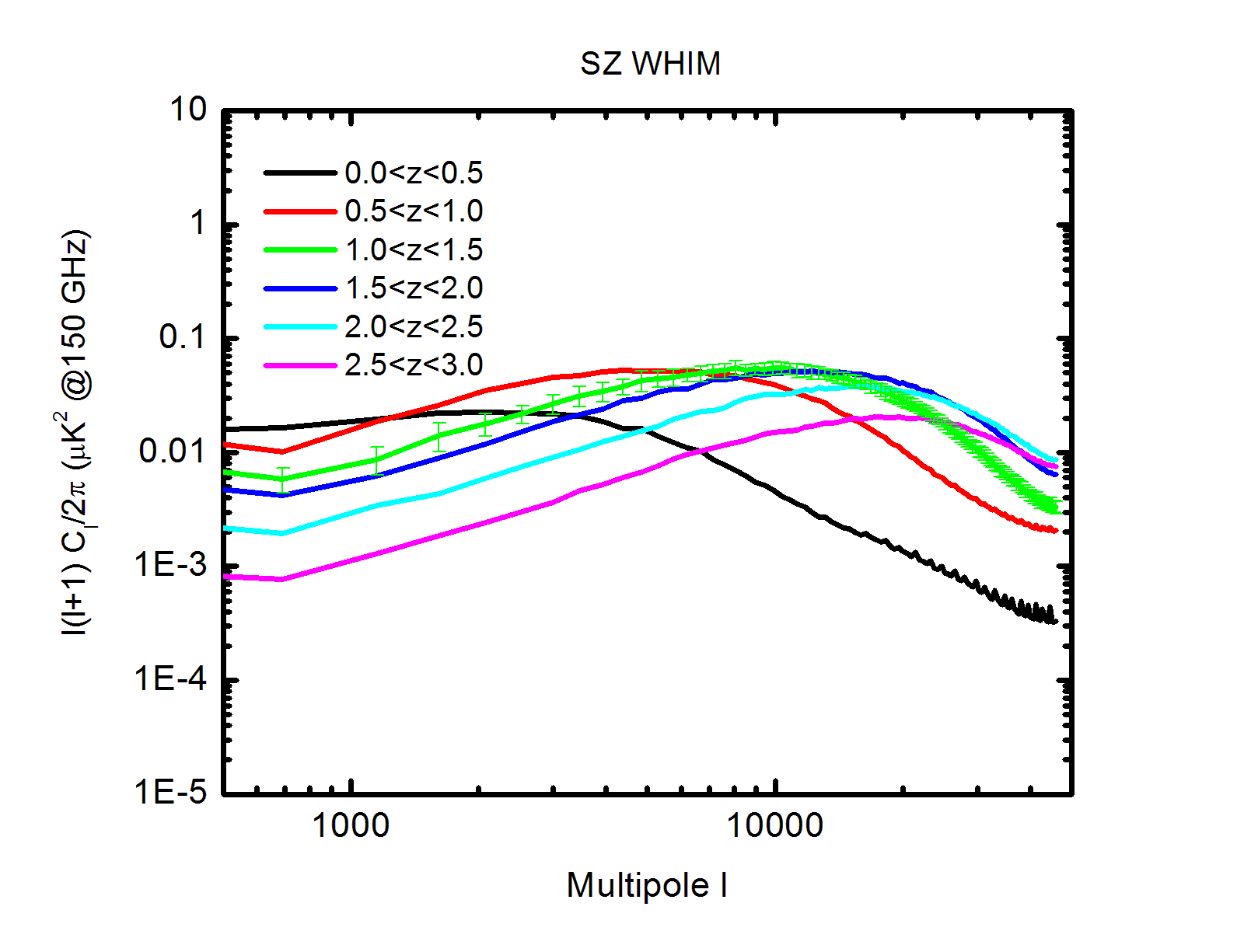}
\caption{Contribution to all-gas (\emph{top}) and WHIM (\emph{bottom}) SZ (at 150~GHz) at equal redshift intervals. Notice that the WHIM and all-gas plots have the same scale for an easier and direct comparison. The figures show that the correlation signal is larger at small redshift. The WHIM signal, though smaller than the all-gas signal, is significantly different from zero. The maxima of the \PS curves are at lower poles for lower redshifts. For sake of clarity we show the variance only for the [$1.0<z<1.5$] interval, the variance of the other signals, however, is comparable with that of the [$1.0<z<1.5$] interval. Color version on-line only.
\label{A_SZ_zs}}
\end{figure*}

In figure~\ref{A_SZ_zs} we show the WHIM and all-gas SZ redshift evolution. In general the analysis of the redshift evolution of the SZ proved to be more difficult than for X-rays, in particular for IGM at redshift below $z=1.0$. The all-gas signal, in fact, shows an extremely high variance and huge clusters completely dominate the statistics as in some cases a single cluster can cover almost 30\% of the map. In X-rays the same structures look much smaller since X-ray emission depends on $\rho^2$ and not on $\rho$ and therefore we probe only the denser region of the objects (ending in a smaller impact on the properties of the average map). This large variability for the SZ prevented us from studying the redshift evolution using smaller redshift intervals, as we needed large statistics to make up for sample variance.

Looking at the plots in figure~\ref{A_SZ_zs}, we see that both the amplitude and the shape of the SZ correlation signal evolve with redshift. The AcF shows that the amplitude is smaller at early time and that the slope becomes steeper, suggesting  smaller characteristic angles. It is interesting to note that the change in amplitude is not so dramatic as in X-rays: in particular, for the WHIM, the signal at $z=3$ is $\sim10\%$ of the signal at $z<1$. In the \PS\ there are significant changes compared to the X-rays. The peaks for the IGM signal are now in the multipole range between $l=3000$ and $l=30000$ (increasing with distance), again because objects in the SZ are larger than their X-ray counterpart. The peaks for the WHIM are located at larger scales, as well, ranging from $l\sim2500$ to $l\sim25000$. The plot also shows that, although the contribution from high redshift WHIM is marginal, it is however the dominant component at $l>30000$: when looking at the smallest scales we are probing almost exclusively the WHIM at $z>1.5$, when it was still in the early stage of its evolution. 

The evolution of the WHIM SZ signal is strictly related to the WHIM formation history. The amount of matter in the WHIM phase, in fact, has been negligible before $z=3.0$ and only after $z=2.0$ it has become a significant fraction of the whole gas. As a result, the WHIM has been contributing to the SZ effect only at a relatively recent stage. 

The IGM SZ signal follows a similar pattern, but the mechanism is different. The density distribution of the gas, in fact, evolves with time, going from an almost uniform distribution of low density gas at high redshift to the patchy distribution with high density regions at present time. As we have already observed, the SZ effect depends on density, and it is the higher fraction of high density gas that gives a stronger signal at low redshift. 

\subsection{Redshift evolution of the cross-correlation signal}
\label{XSZ_corr_evo}

\begin{figure*}
\plottwo{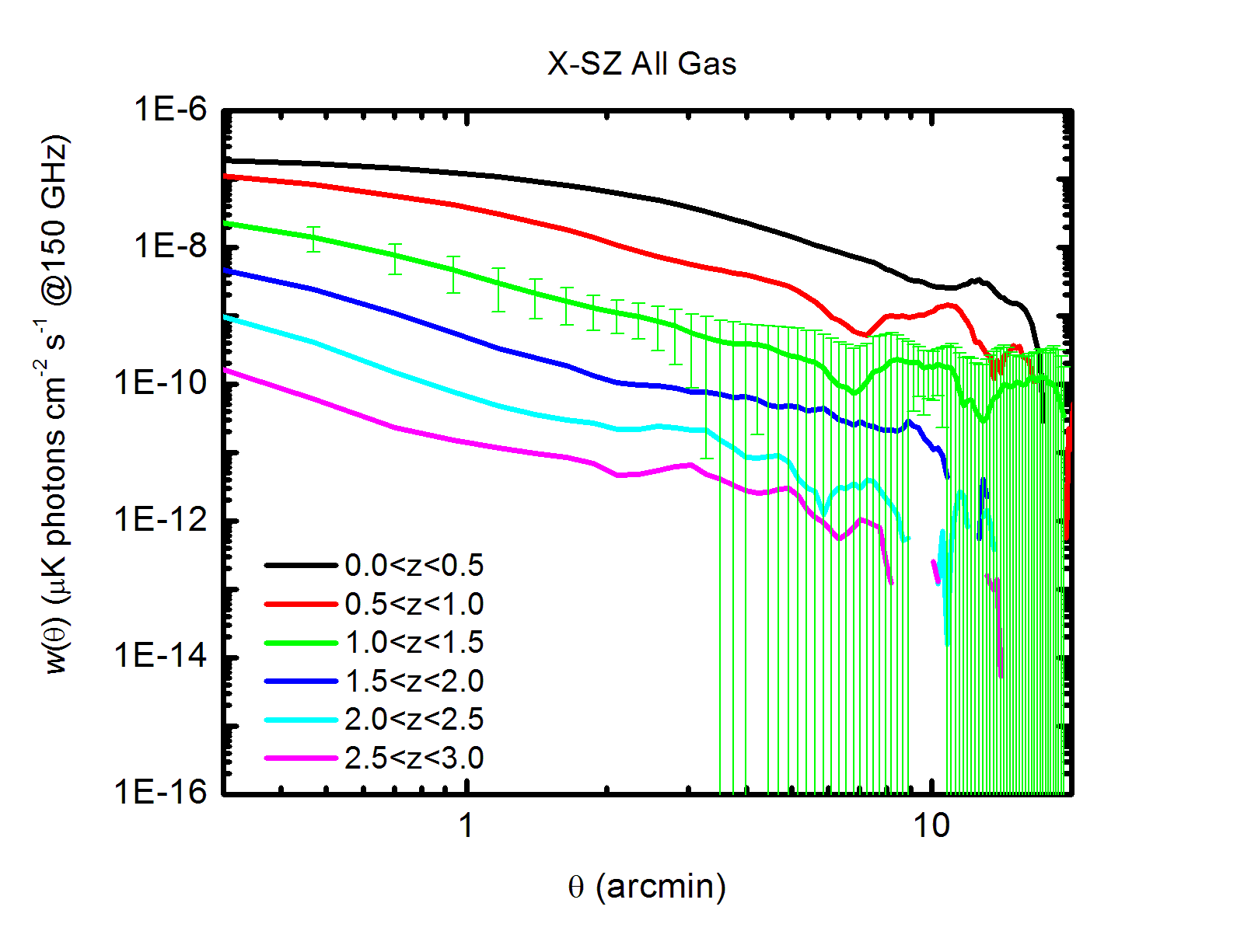}{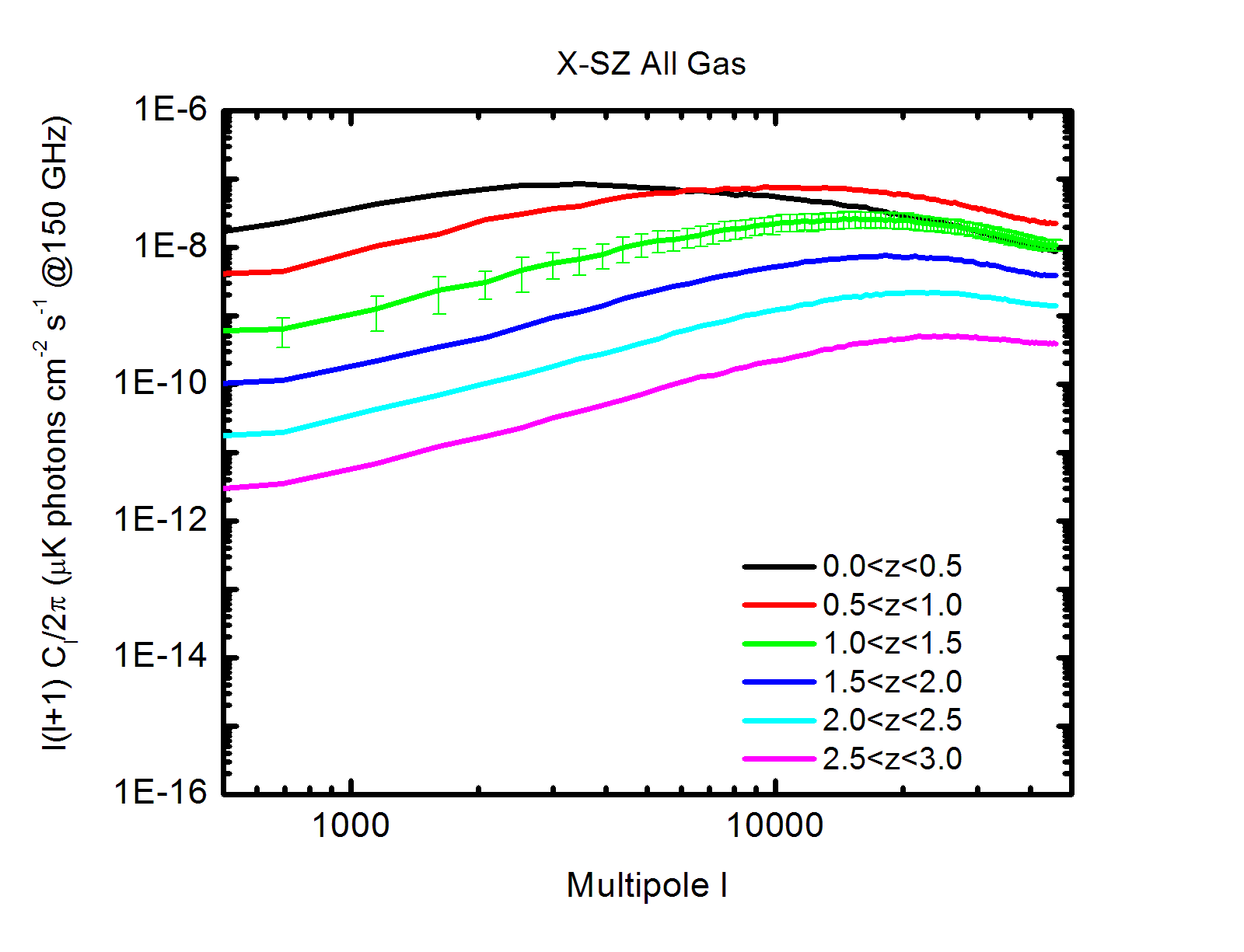}
\plottwo{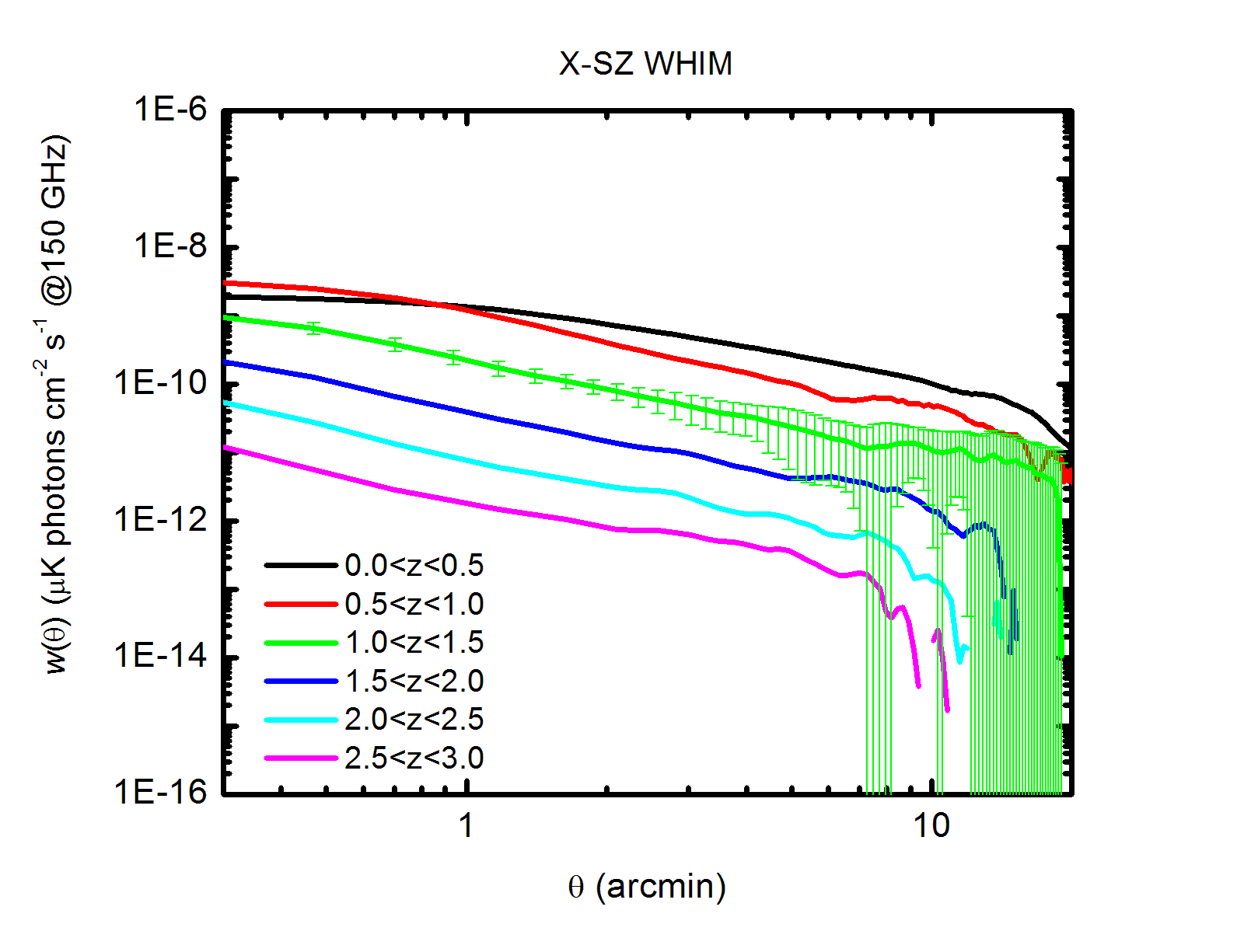}{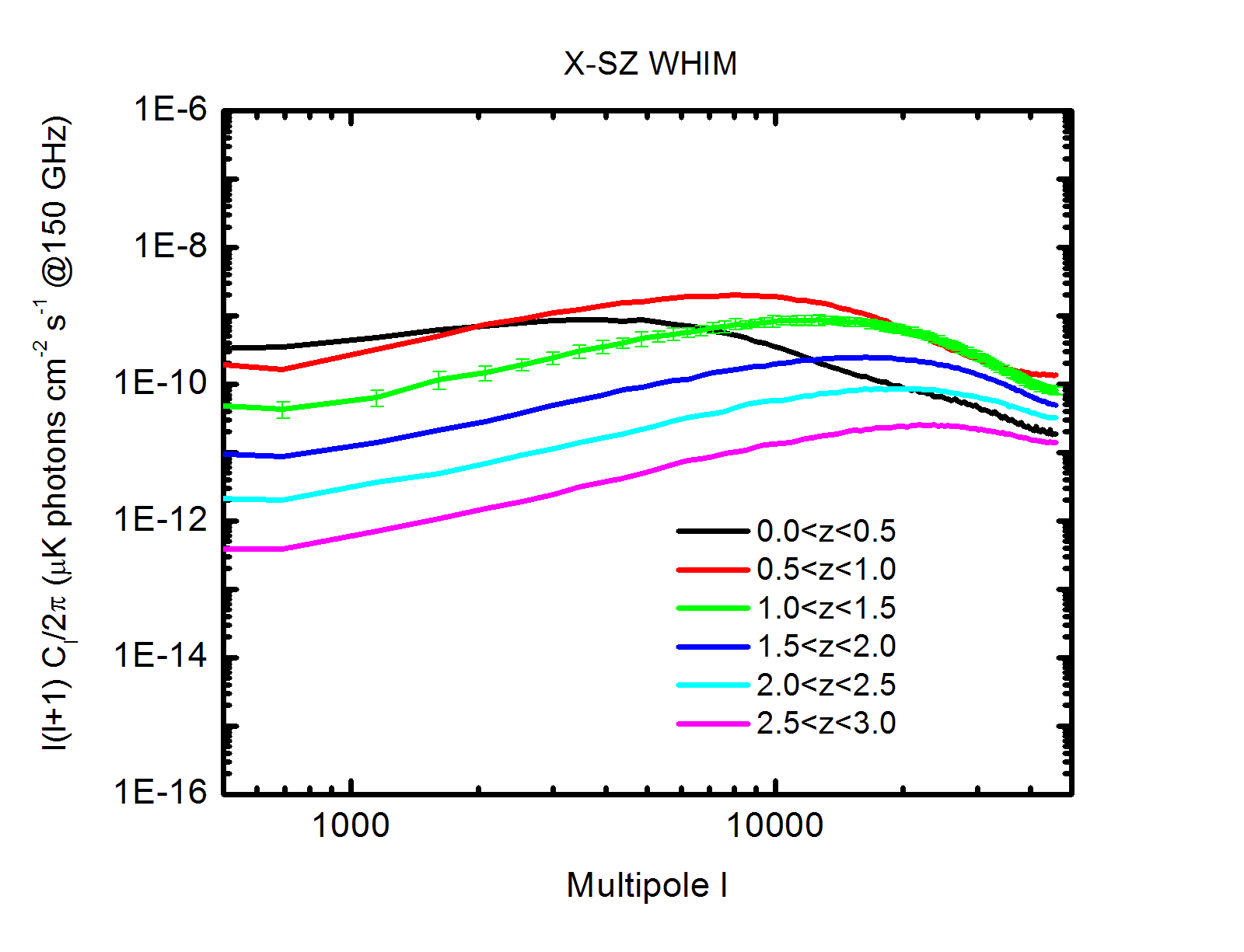}
\caption{Correlated contribution to X-rays and SZ due to all the gas (\emph{top}) and the WHIM (\emph{bottom}) at equal redshift intervals. The figures show that the correlation signal is larger at small redshift. The WHIM signal, though smaller than the all-gas signal, is significantly different from zero. The maxima of the \PS curves are at lower poles for lower redshifts. For sake of clarity we show the variance only for the [$1.0<z<1.5$] interval, the variance of the other signals, however, is comparable with that of the [$1.0<z<1.5$] interval. Notice that the WHIM and all-gas plots have the same scale for an easier and direct comparison. Color version on-line only.
\label{A_XSZ_zs}}
\end{figure*}

In figure~\ref{A_XSZ_zs} we show how the cross-correlation between X-rays and SZ evolves with redshift. The amplitude of the WHIM signal is low at early time and grows by about two orders of magnitude until $z=1$, after which it remains fairly constant. Similarly to the X-ray and SZ \PS, also the cross-power peak shifts towards higher multipoles as the redshift increases, even when the total amplitude does not show significant variations ($z<1$). The position of the peaks, however, is different from the one we find at the same redshifts for the X-ray emission or the SZ. Similarly to the SZ (lower right panel of figure~\ref{A_SZ_zs}), also the cross-power offers a tool to directly probe the moderate high redshift WHIM: at $l>20000$, in fact, half of the signal comes from $1.0<z<2.5$ (while the WHIM at $z>2.5$ contributes only for a few percent). The evolution of the IGM cross-correlation signal is very similar to that of the WHIM: it grows from early time until $z=1$, then it remains relatively stable. The cross-spectrum peak, as well, shifts with time throughout all the multipole range. Unlike the WHIM, however, the all-gas signal is dominated by the lower redshift ($z<1.5$) gas at all multipoles.

\section{Conclusions}
\label{conclusion}
In this work we have used hydrodynamical simulations to generate maps of the X-ray emission (in the [$0.4-0.6$]~keV band) and temperature fluctuations due to thermal SZ for the IGM and the WHIM with the aim of investigating the properties of large scale structure and identifying new method to study the elusive signal of the WHIM. 

The X-ray emission from the IGM turns out to be too high compared to observational data, but this is a well-known problem for our simulation and it is restricted to the densest regions \citep{Borgani04}. The emission from the WHIM, on the other hand, is well within the observational constraints both in terms of average emission and in terms of angular correlation. The signal of the WHIM has a characteristic angle of a few arcminutes or, in terms of \PS, it peaks at $l\sim10000$. The bulk of the WHIM emission comes from distances at $z<1$ and becomes negligible at $z\gtrsim1.5$, this is due to the time evolution of the WHIM and to spectral selection effects. Similarly, the contribution to the AcF/PS signal comes mostly from $z<1$, although the shape of the correlation signal changes with distance, favoring smaller angles for more distant objects. The X-ray signal from the WHIM is anti-correlated with the IGM emission at small scales ($l>25000$), the presence of WHIM emission excludes the non-WHIM emission (and vice-versa).
	
The IGM SZ \PS\ predicted by our simulation is in good agreement with other simulations and with data measured by \Planck, \ACT, and \SPT. Unfortunately, the magnitude of the WHIM signal is about fifty times smaller than that of the IGM (a difference smaller than for X-rays) and is too small to be identified with present time detectors. The WHIM SZ signal comes almost uniformly from matter at distances up to $z=2$ but, although it becomes much smaller, it is not completely negligible even beyond $z=3$. Also the amplitude and the characteristic angle of the correlation signal are smaller at earlier times, although the characteristic angle is larger than the corresponding angle in X-rays. Unlike the X-ray signal, however, at small scales ($l>25000$) the SZ \PS\ probes almost exclusively the early time WHIM ($z>1.5$). In the SZ the WHIM and non-WHIM signals do not show anti-correlation (as it happens instead in X-rays): apparently, the fact that the SZ depends (almost) linearly on density smooths the transition between WHIM and non-WHIM.

The X-ray emission and the SZ depend on different physics: SZ scales as the density while the X-ray emission goes as $\rho^2$. This reflects in the different characteristic angles between X-ray and SZ maps: the SZ signal is more uniform over a large portion of an object while the X-rays come mostly from its densest parts, usually located at the core of clusters and of filaments. 

The 
correlation between X-ray and SZ maps is largely dominated by the non-WHIM component. There are however several considerations that make the analysis of the cross-correlation between X-rays and SZ an interesting tool to study the WHIM. 

\begin{enumerate}
	\item The cross-correlation signal is the sum of the WHIM cross-correlation, the non-WHIM cross-correlation, and the mixed terms. Since the WHIM cross-correlation 	term is the smallest component, it is necessary to remove all the other terms. The non-WHIM cross-term is by far the strongest and can be easily reduced by 	removing all the sources detected in X-rays. The term from WHIM SZ and non-WHIM X-ray, as well, is about one order of magnitude stronger than the WHIM signal, and it is strongly reduced when we remove the X-ray sources. The remaining mixed term, the one obtained with non-WHIM SZ and WHIM X-ray has amplitude very similar to that of the WHIM cross-correlation (at least at scales smaller than a few arcminutes), the mask adopted to remove clusters and point sources in the X-ray map should be already enough to reduce it. The technical feasibility of this approach, however, will be investigated in future work.
	\item The cross-correlation WHIM signal at the scale of a few arcminutes is dominated by the gas within $z<1.5$, thus complementing the information that can be extracted from the X-ray AcF.
	\item The cross-correlation WHIM signal at very small scales is dominated by distant ($z>1.5$) gas, giving us a unique tool to study the properties of the WHIM in its early stage, where the analysis of the X-ray maps alone, although easier, limits our investigation to the WHIM at $z<1.0$. 
\end{enumerate}

\acknowledgments

This work has been supported by NASA grant NNX11AF80G. The authors would like to thank Stefano Borgani for the access to his hydrodynamical simulations and Enzo Branchini for the useful discussion and suggestions.

\end{document}